\documentclass{jfm}

\usepackage{graphicx}
\usepackage{natbib}

\newcommand{\rd}{\mathrm{d}}
\newcommand{\rbar}{\bar{R}}

\newcommand{\rmin}{R_{\mathrm{min}}}

\ifCUPmtlplainloaded \else
  \checkfont{eurm10}
  \iffontfound
    \IfFileExists{upmath.sty}
      {\typeout{^^JFound AMS Euler Roman fonts on the system,
                   using the 'upmath' package.^^J}%
       \usepackage{upmath}}
      {\typeout{^^JFound AMS Euler Roman fonts on the system, but you
                   dont seem to have the}%
       \typeout{'upmath' package installed. JFM.cls can take advantage
                 of these fonts,^^Jif you use 'upmath' package.^^J}%
      }
  \else
  \fi
\fi

\ifCUPmtlplainloaded \else
  \checkfont{msam10}
  \iffontfound
    \IfFileExists{amssymb.sty}
      {\typeout{^^JFound AMS Symbol fonts on the system, using the
                'amssymb' package.^^J}%
       \usepackage{amssymb}%
         \let\leq=\leqslant
         
      }{}
  \fi
\fi

\ifCUPmtlplainloaded \else
  \IfFileExists{amsbsy.sty}
    {\typeout{^^JFound the 'amsbsy' package on the system, using it.^^J}%
     \usepackage{amsbsy}}
    {\providecommand\boldsymbol[1]{\mbox{\boldmath $##1$}}}
\fi

\newsavebox{\astrutbox}
\sbox{\astrutbox}{\rule[-5pt]{0pt}{20pt}}

\title[Guidelines for authors]{Vibration and Nonlinear Resonance in the Break-up of an Underwater Bubble}

\author[L. Lai, N. C. Keim, K. Fezzaa, W. W. Zhang and S. R. Nagel]%
{Lipeng Lai$^1$%
  \thanks{Email address for correspondence: lplai@uchicago.edu}, Nathan C. Keim$^1$, Kamel Fezzaa$^2$, Wendy W. Zhang$^1$ and Sidney R. Nagel$^1$}

\affiliation{$^1$Physics Department and the James Franck Institute, University of Chicago, 929 E. 57th St., Chicago IL 60637, USA\\
$^2$X-ray Science Division, Argonne National Laboratory, 9700 S. Cass Ave., Argonne IL 60439, USA}

\pubyear{}
\volume{}
\pagerange{}
\begin{document}

\maketitle

\begin{abstract}
We use high-speed X-ray phase-contrast imaging, weakly nonlinear analysis and boundary integral simulations to characterize the final stage of underwater bubble break-up.  The X-ray imaging study shows that an initial azimuthal perturbation to the shape of the bubble neck  gives rise to oscillations that increasingly distort the cross-section shape.  These oscillations terminate in a pinch-off where the bubble surface develops concave regions that contact similar to what occurs when two liquid drops coalesce. We also present a weakly nonlinear analysis that shows that this coalescence-like mode of pinch-off occurs when the initial shape oscillation interferes constructively with the higher harmonics it generates and thus reinforce each other's effects in bringing about bubble break-up. Finally we present numerical results that confirm the weakly nonlinear analysis scenario as well as provide insight into observed shape reversals.  They demonstrate that when the oscillations interfere destructively,  a qualitatively different mode of pinch-off results where the cross-section profile of the bubble neck develops sharply-curved regions.  
\end{abstract}


\section{Introduction}

To a large extent, our understanding of how nonlinearity organizes continuous fields in motion are informed by two processes:  nonlinear resonance and singularity formation. Nonlinear resonance generates a wealth of complex motion from a few, initially simple ingredients \citep{cross93}.  In contrast, singularity formation often causes an initially complicated state to evolve into a simple form.  As stresses in the neighborhood of the singularity diverge, the dynamics becomes dominated by the presence of the singularity. In the most extreme cases, the singularity dynamics approaches a universal form, one independent of initial or boundary conditions \citep{shi94,egg97,barenblatt96}.  

At first sight, these two processes seem confined to mutually exclusive regimes.  The full effect of nonlinear resonance requires many iterations and a spatially-extended system.  Finite-time singularity formation occurs when the characteristic length- and time-scales go to zero.  However, we do know many examples where nonlinear resonance gives rise to the formation of a singularity.  In wave breaking, the generation and amplification of higher harmonics by nonlinear resonance among the different modes are so efficient that a finite-time singularity results \citep{kuznetsov94,dyachenko96}.  The onset of a period-doubling cascade as a nonlinear system's parameter is tuned towards a critical value, such as occurs in Rayleigh-B\'enard convection or in the dripping of water drops from a faucet, are other examples \citep{predrag89,ott}. In contrast, few phenomena corresponding to the obverse scenario, singularity formation being perturbed, or even cut-off, by nonlinear resonance, have been identified.  Here we present evidence from experiments, theory and simulation that the familiar phenomenon of a bubble breaking into several bubbles while underwater provides such an example.  

Previous studies have established that the cylindrically-symmetric underwater bubble break-up is dynamically unstable \citep{keim06,schmidt09}.  Minute azimuthal variations in the initial state of the dynamics give rise to standing waves whose effects become more pronounced as pinch-off approaches.  Intriguingly, when the size of the azimuthal perturbation is small, the cylindrically-symmetric singularity dynamics does not select among the different modes present by amplifying some modes and diminishing others.  Instead of this behavior, each Fourier mode excites a shape oscillation whose amplitude remains constant over time, thereby encoding a memory of the initial state.  This absence of selection contrasts sharply with other dynamically unstable singularities, such as the break-up of a glycerine-water drop in air, or the pinch-off of a viscous oil thread \citep{shi94,brenner96}.  In these examples, the instability associated with the pinch-off singularity is dominated by a single growing mode corresponding to a shift in the onset time and location of the singularity.  In underwater bubble break-up, the preservation of the amplitudes of the initial perturbation modes  continues until the dynamics is strongly distorted from the cylindrically-symmetric one.  As a result, the final form of the bubble break-up dynamics is shaped by nonlinear resonance among the different shape-oscillation modes that have been excited.

Here we use high-speed X-ray imaging to visualize the three-dimensional evolution that characterizes the final stage of underwater bubble break-up and its aftermath.  These results confirm that the nonlinear distortions from shape vibrations become important in the final stage of break-up. High-speed visible-light photography studies \citep{burton05,keim06,bergmann06, thoroddsen07} are not able to provide information about the surface evolution in this regime because the air-water surface becomes re-entrant, thus obscuring key features. We also use weakly nonlinear analysis and boundary integral simulations to assess how the final dynamics varies as a function of the initial perturbation.  The nonlinear resonance between an initial perturbation and the first higher harmonic it generates gives rise to two different types of break-up modes.  When the two distinct shape oscillations act in concert to bring out a topology change, a coalescence-like mode of pinch-off results.  In contrast, when the two shape vibrations interfere destructively so that one mode opposes another's action, many higher harmonic modes are amplified.  The cross-section of the bubble neck becomes highly elongated with sharply curved ends, corresponding to a cusp-like mode of pinch-off. 

\section{Background}

\begin{figure}
\centerline{\includegraphics[scale=0.4]{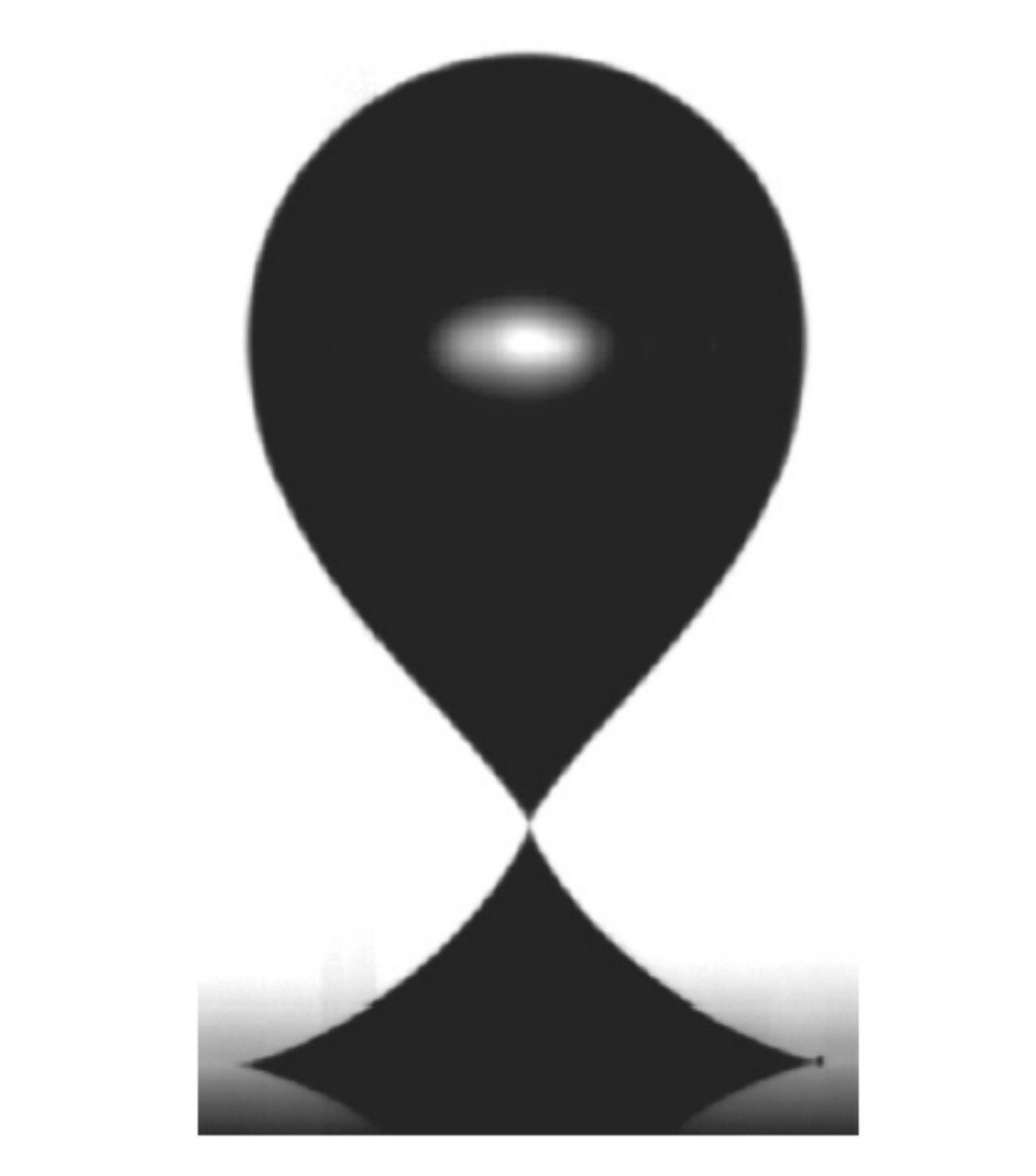}}
\caption{Experiment: underwater bubble pinch-off.  The bubble is blown from a $6$-mm diameter circular hole cut in a plate that was carefully leveled.  The plate reflects the gas-liquid interface.  The bright spot in the dark bubble is due to back-lighting. Image reproduced from Keim \citep{keim11}.}
\label{fig:pinchoff}
\end{figure}

To describe more specifically how the cylindrically symmetric pinch-off is pre-empted by the dynamics excited by initial azimuthal asymmetries, we focus on what happens in the quasi-static bubble-release experiment (figure \ref{fig:pinchoff}): an underwater bubble is made by allowing gas to leak very slowly into a tank of water through an aperture in a rigid plate.  When the bubble is sufficiently large, buoyancy overcomes surface tension and the bubble pinches off.  If the pinch-off dynamics remained cylindrically-symmetric throughout, then a finite-time singularity in the continuum description forms. To see this, firstly it is noted that high-speed photography measurements show that both $\rmin$--- the minimum neck radius, and $R_{\mathrm{ax}}$--- the radius of curvature associated with how the surface curves along the length of the bubble neck, decrease approximately as $(t_* - t)^{\alpha}$ with $\alpha \approx 0.56$ as the moment of pinch-off $t_*$ approaches \citep{burton05, keim06, thoroddsen07} (The power-law decrease in $\rmin$ has a larger numerical prefactor than the decrease in $R_{\mathrm{ax}}$. Otherwise the bubble neck would evolve towards a catenoid, and therefore, stable configuration.). 

The value of the scaling exponent $\alpha$ is associated with a two-dimensional flow dynamics, one in which the momentum balances at different heights along the bubble neck axis are essentially decoupled \citep{longuet91,oguz93}.  While pinch-off involves motion of both the liquid in the exterior and the gas in the interior of the bubble, the liquid has a much larger density than the gas and therefore dominates the momentum balance.  The contraction of the bubble neck requires that some amount of the liquid in the exterior is accelerated inwards along with the interface.  If the minimum radius of the bubble neck at time $t$ is $\rmin(t)$, the amount of water of density $\rho$ that needs to move inwards to fill the circular hole previously occupied by air is approximately $\rho \pi \rmin^2(t)$.  As a result, the kinetic energy per unit length of water motion produced by break-up, $\rho \pi \rmin^2(t) (\rd {\rmin} / \rd t)^2$, decreases as  $(t_*-t)^{4 \alpha -2}$ or $(t_* - t)^{0.24}$.  This is a weak decrease.  For comparison, the kinetic energy in the neighborhood of break-up decreases as $(t_* - t)^{4/3}$ in the singularity dynamics associated with the pinch-off of a water drop in air.   We therefore view the underwater pinch-off process as an energy focusing process.  The cylindrically-symmetric dynamics concentrates a finite amount of kinetic energy into the vanishingly small break-up region (Actually, as often obtains in nearly two-dimensional evolution, the power law form is modified by a  logarithmically slow evolution in the value of $\alpha$ \citep{eggers07,gekle09b}. In practice, the power-law divergence is an excellent description.  This is because even for the largest stable bubble we can make on earth, the inertial pinch-off dynamics begin with the bubble neck around $1$ mm across in its thinnest section, while the continuum description breaks down below $1$ nm. The divergent behavior associated with pinch-off obtains over at most $6$ decades of length-contraction, too narrow a range for a logarithmic variation to produce a significant effect.).   

One consequence of this energy focusing is that the divergence in the Laplace pressure is no longer the fastest growing singularity. To see this, if we compare the magnitude of the Laplace pressure $\sigma \kappa$, where $\sigma$ is the surface tension and $\kappa =  (1/\rmin + 1/R_{\mathrm{ax}})/2$ is the mean curvature of the neck, versus the Bernoulli pressure $\rho (\rd \rmin / \rd t)^2/2$, the Laplace pressure diverges as $(t_* - t)^{-0.56}$.  This is slower than the $(t_*-t)^{-0.88}$ divergence in the Bernoulli pressure.  We are used to thinking of surface tension effects as negligible when the characteristic length scale is large but important when the length scale is small.  The break-up of an underwater bubble is an exception to this rule.  The momentum balance governing the evolution of the bubble shape in a quasi-static release experiment (figure \ref{fig:pinchoff}) begins by being dominated by surface tension but ends, at the smallest length-scale near break-up, being dominated by inertia.  From this perspective, underwater bubble break-up dynamics does not belong with surface tension-driven free-surface singularities that characterize other fluid drop break-ups, such as the dripping of a water drop off the end of a faucet or the snap-off of an oil drop in water \citep{egg97}.   It is more natural in fact to classify the cylindrically-symmetric bubble pinch-off dynamics with focusing singularities that show up in models of cavity collapse, shock implosion, or supernova formation \citep{whitham57,rygg08,plewa04,maeda08}.
\begin{figure}
\centerline{\includegraphics[scale=0.6]{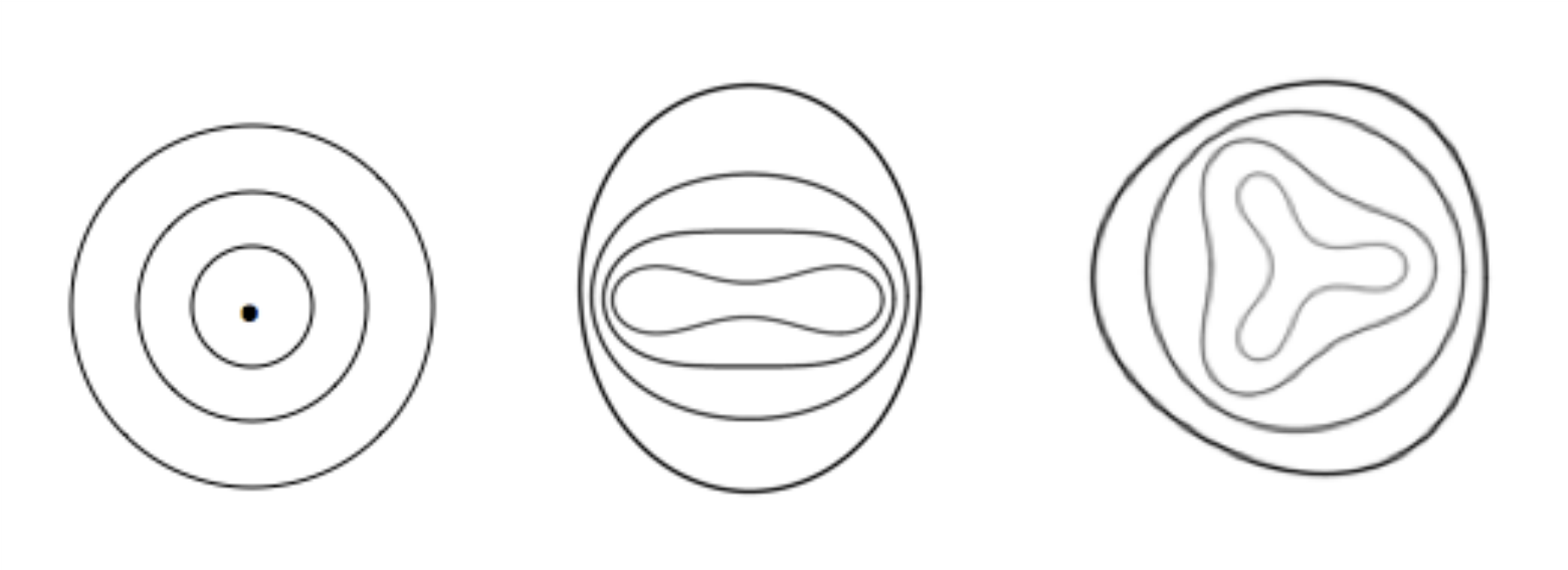}}
\caption{Schematic illustrating linear stability dynamics described by equations (\ref{linstab_amp}) and (\ref{linstab_phase}).  From left to right, we show what results when the initial bubble cross-section is circular, distorted by a single $n=2$ Fourier mode, or a single $n=3$ Fourier mode.  Over time, the cross-section contracts and distorts into shapes preserving the original symmetry of the imposed perturbation.}
\label{fig:azimuthal}
\end{figure}

\begin{figure}
\centerline{\includegraphics[width=10cm]{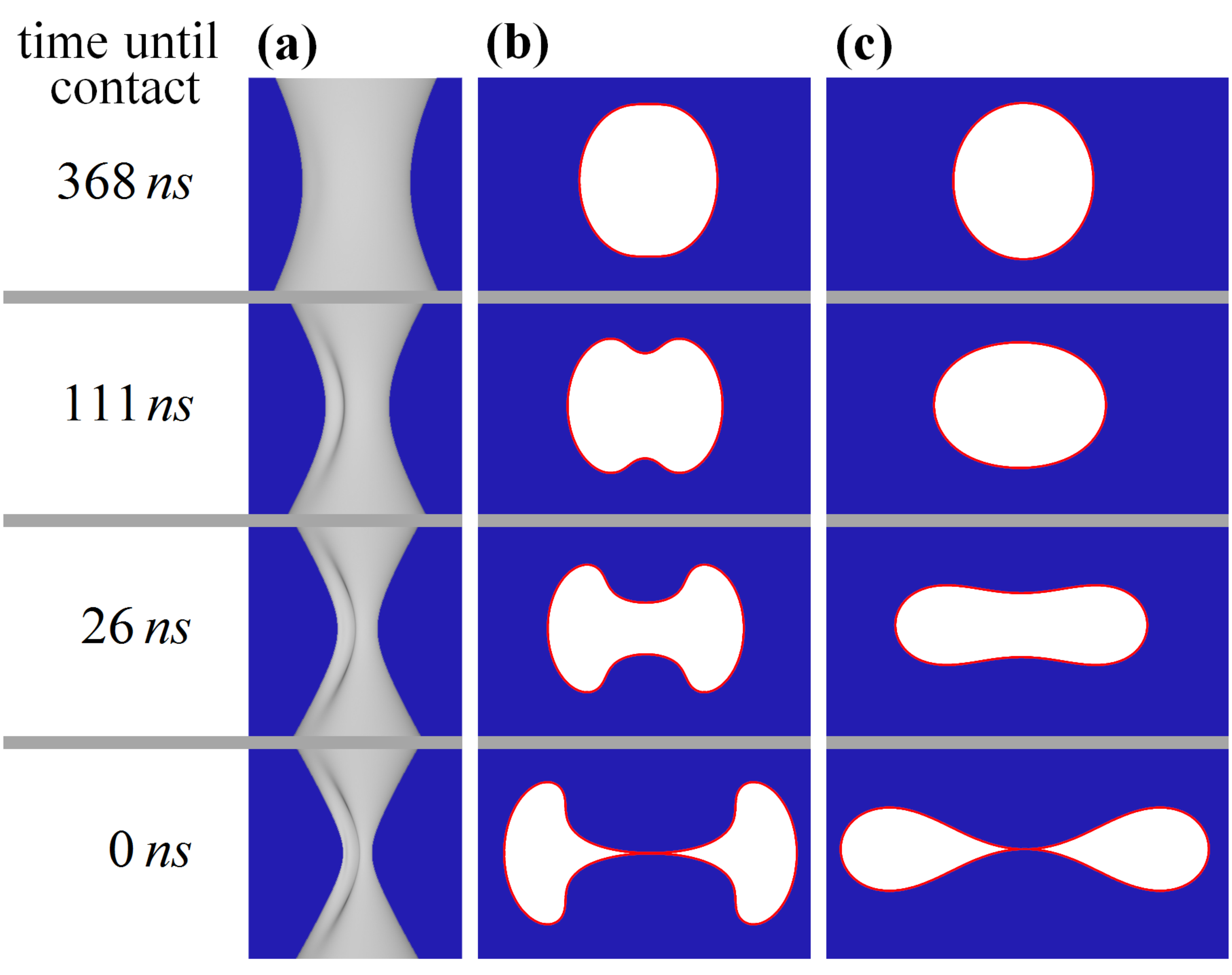}}
\caption{Simulated and predicted neck evolution from a model that assumes the pinch-off dynamics is dominated by radial influx of water and is therefore effectively two-dimensional. The left-most column gives the time until break-up. Column (a) displays the three-dimensional neck shapes generated by stacking simulation results from the two-dimensional model vertically. Column (b) displays the evolution of the cross-section at the height where the bubble neck radius has its minimum value. Column (c) displays the cross-section evolution at the neck minimum predicted by linear stability analysis using approximately canonical coordinates $\cal R$, $\cal V$ (see Section \ref{subsec:RV} for more details). The profiles show qualitative agreement with the full simulation results (column (b)). In order to display clearly how the cross-section shape distorts over time, we have rescaled the successive profiles in columns (b) and (c) by the average cross-section radius. The short time scales in the left-most column underscore the power of the simulation and the analytical model in revealing the dynamics inaccessible to experiments. Please see Section \ref{sec:prob} and \ref{sec:num} for more details.}
\label{fig:2dmodel}
\end{figure}

The increasing irrelevance of surface tension causes the cylindrically-symmetric underwater bubble break-up to become dynamically unstable.  The $O(1)$, symmetric pinch-off dynamics allows any minute azimuthal perturbations present to retain their absolute amplitudes over time. Specifically, while the perturbation amplitude is small, the shape $S(\theta, \rbar(t))$ of the bubble neck cross-section with average radius $\rbar(t)$ has the form 
\begin{equation}
S(\theta, \rbar(t)) = \rbar(t) + \Sigma_n A_n \cos[ \phi_n(\rbar(t))] \cos (n\theta)
\label{linstab_amp}
\end{equation}
where $\bar{R}$ defines the radius of a circle having the same area enclosed as the cross-section (to the first order of $A_n/\bar{R}$). $\Sigma_n$ denotes summation over all the Fourier modes initially present in the azimuthal perturbation, and $A_n$ is the absolute amplitude of each mode and remains constant over time. The phase of each shape oscillation $\phi_n$ winds up logarithmically as the average radius of the cross-section decreases \citep{schmidt09}, 
\begin{equation}
\phi_n(\rbar) = \sqrt{n-1} \ln[ \rbar_0 / \rbar(t)] + \Omega_n
\label{linstab_phase}
\end{equation}
where $\rbar_0$ is the average radius of the cross-section at $t=0$ and $\Omega_n$ is the initial value of the phase variable. Since the minimum radius of the bubble neck tends towards $0$, azimuthal perturbations, however small initially, grow in relative importance.  The cross-section of the bubble neck near the minimum becomes increasingly distorted from a circular shape.  Figure \ref{fig:azimuthal} provides a few examples of the kinds of distortion that obtains when the initial perturbation is simply a single Fourier mode.  The perturbation gives rise to a shape vibration of the bubble surface, with concomitant vibrations in the velocity and pressure fields in the exterior.  As time goes on, the vibration starts to interact strongly with the symmetric dynamics and to generate higher harmonic modes.  Eventually the relative amplitude of the vibration becomes so large that it causes the cross-section profile of the bubble neck to curve inwards and contact at a point (figure \ref{fig:2dmodel}), thus pre-empting the cylindrically-symmetric pinch-off. This effect is most exaggerated at the neck minimum, where the first contact occurs, both because the relative amplitude of the perturbation is the largest there and because the phase of relevant perturbation changes most rapidly as $\rbar$ approaches $0$.  Thus, even in the simplest situation where the shape vibration does not complete a half cycle and therefore change its orientation (figure \ref{fig:2dmodel}), the shape of the bubble neck is fully three-dimensional at pinch-off.  The interface develops creases that emanate from the contact point. The experimental data in figure \ref{fig:xraymemory} shows the more complex outcome that is obtained when higher modes interfere with the fundamental modes destructively. The presence of the higher harmonic modes becomes evident when one compares the cross-section shape calculated in a full simulation (column (b) in figure \ref{fig:2dmodel}) against extrapolations of linear stability dynamics down to the moment of contact (column (c) in figure \ref{fig:2dmodel}).  While the shape evolutions on the whole agree, the higher harmonic modes cause the cross-section shape to become more curved than the shapes produced by extrapolations of the linear stability dynamics. 

Turitsyn, Lai and Zhang conducted a systematic study of the final outcome as a function of the initial mode amplitude when the initial mode is $n=2$ \citep{turitsyn09}.  They found that the final break-up dynamics changes abruptly as  the orientation of the coalescence plane switches discontinuously between the two allowed values.   Our study is motivated by the suggestion from their analysis that the final break-up dynamics displays heightened sensitivity to initial conditions in the narrow intervals where the orientation of the coalescence-mode break-up changes.  We use high-speed X-ray photography to image directly the final stage of bubble break-up (Section \ref{sec:exp}). The results confirm findings from the numerical simulations that the shape evolution in the final moments of break-up produces re-entrant cross-section profiles.  In addition, the imaging measurements reveal that slight tilts can profoundly alter the jet formation after bubble break-up.  In Section \ref{sec:prob} we present a mathematical model of the break-up dynamics, a review of linear stability results and a weakly nonlinear analysis showing that destructive interference between the original perturbation mode and the first higher harmonic it generates heightens the effect of nonlinearity. In Section \ref{sec:num}, we present results from boundary integral simulations, which show that the onset of destructive interference is accompanied by a qualitatively different type of the break-up. We also provide a possible explanation for the X-ray observations in figure \ref{fig:xraymemory} based on our simulation results. We discuss questions raised by our results in Section \ref{sec:discussion} and conclude with a summary of key results in Section \ref{sec:conclusion}.
\newpage

\section{Experiment}\label{sec:exp}
\subsection{Experimental Methods}

\begin{figure}
\centerline{\includegraphics[width=3in]{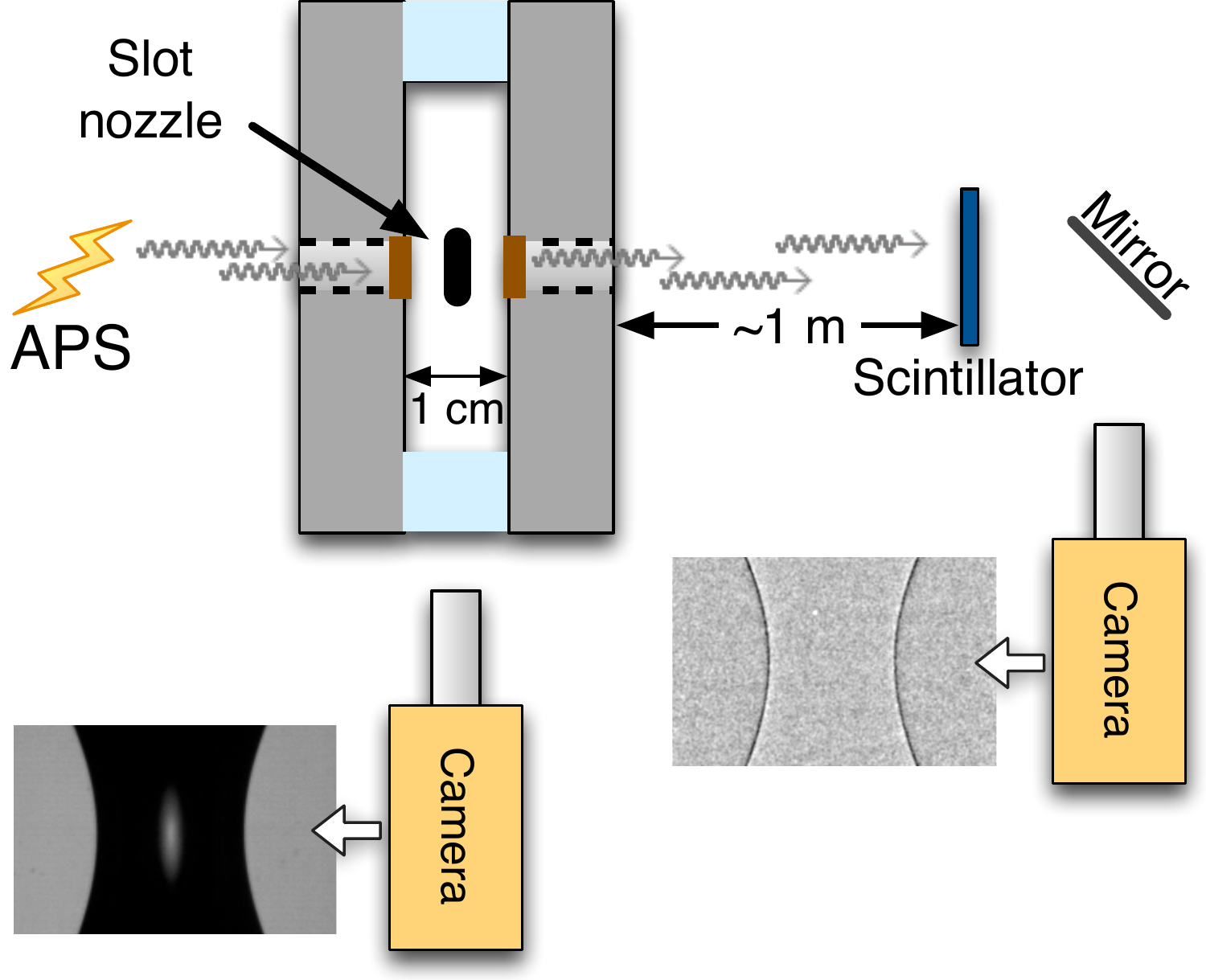}} 
\caption{Top view of the synchrotron apparatus for imaging pinch-off with X-rays. Collimated X-rays from the Advanced Photon Source (left) interact with the gas-water interface of the bubble at the nozzle, and produce a phase-contrast image on a scintillator. We use two cameras at orthogonal orientations.  One records a high-speed movie of the scintillator while another records the bubble surface evolution in visible light.  Putting the two sets of movies together gives us a fully three-dimensional view of the break-up process while also displaying the formation of re-entrant regions along the bubble surface.  In the schematic, sample frames are shown next to their corresponding cameras. Each line in the phase-contrast image corresponds to X-rays passing tangent to the gas-water interface. 
A slot-shaped nozzle with dimensions $6.3$ $\times$ $1.6$~mm is shown; experiments were also performed with a slot parallel to the X-ray beam, and with a circular nozzle of diameter $6$ mm. 
All X-ray experiments reported here used He gas. 
}
\label{fig:xrayapparatus}
\end{figure}

Experiments that image the asymmetry of pinch-off in visible light must do so primarily by measuring the outer profile of the bubble neck \citep{bergmann06,keim06,schmidt09,keim11}. Thus regions of the neck which are concave, or are internal to the cavity, are generally not observed by these experiments --- in particular, the vertical jets created after pinch-off \citep{gekle09} and the concavities predicted from shape vibrations \citep{keim06, schmidt09, keim11}. Experiments that view the interior of the cavity from above \citep{enriquez12} are also subject to this limitation, as a cavity cross-section shape at one height may be partly occluded by the different cross-section shapes above it.  However, at X-ray wavelengths, the indices of refraction of the gas and water are very close, so that the interface refracts light only slightly, and thus the entire bubble interior may be viewed clearly. To image internal features of the bubble, we performed experiments at the high-speed phase-contrast imaging beamline $32$-ID at the Advanced Photon Source at Argonne National Laboratory \citep{fezzaa08}, using the apparatus shown in figure \ref{fig:xrayapparatus}. 

The phase-contrast technique exploits the slight difference in the indices of X-ray refraction between the two fluids: when a collimated X-ray beam propagates tangentially past a gas-liquid interface, the rays that pass on each side of the interface have slightly different phases, and so form an interference pattern on an image screen some distance away. In these experiments, the pattern consists of a single pair of dark and white fringes that trace out where the interface was tangent to the beam, as shown in the sample X-ray frame in figure \ref{fig:xrayapparatus}. The precise appearance and strength of the fringes has a non-trivial dependence on the details of the interface geometry and the beamline setup, but it is stronger where the interface has a smaller local curvature. Movie frames, recorded here at up to 68,000 frames/s, correspond to single $500$-ns pulses from the synchrotron ring. To remove artifacts in the images due to defects in the scintillator and the Kapton X-ray windows, each pixel in each frame was divided by the mean value of that pixel for the entire movie. Additionally, to ease identification of the fine features in figure \ref{fig:xraymemory}, a bandpass filter from 2 to 40 pixels (4--80 $\mu$m) was applied to each frame with the software ImageJ \citep{imagej}.

\subsection{X-ray observations of shape vibrations}

\begin{figure}
\centerline{\includegraphics[width=12cm]{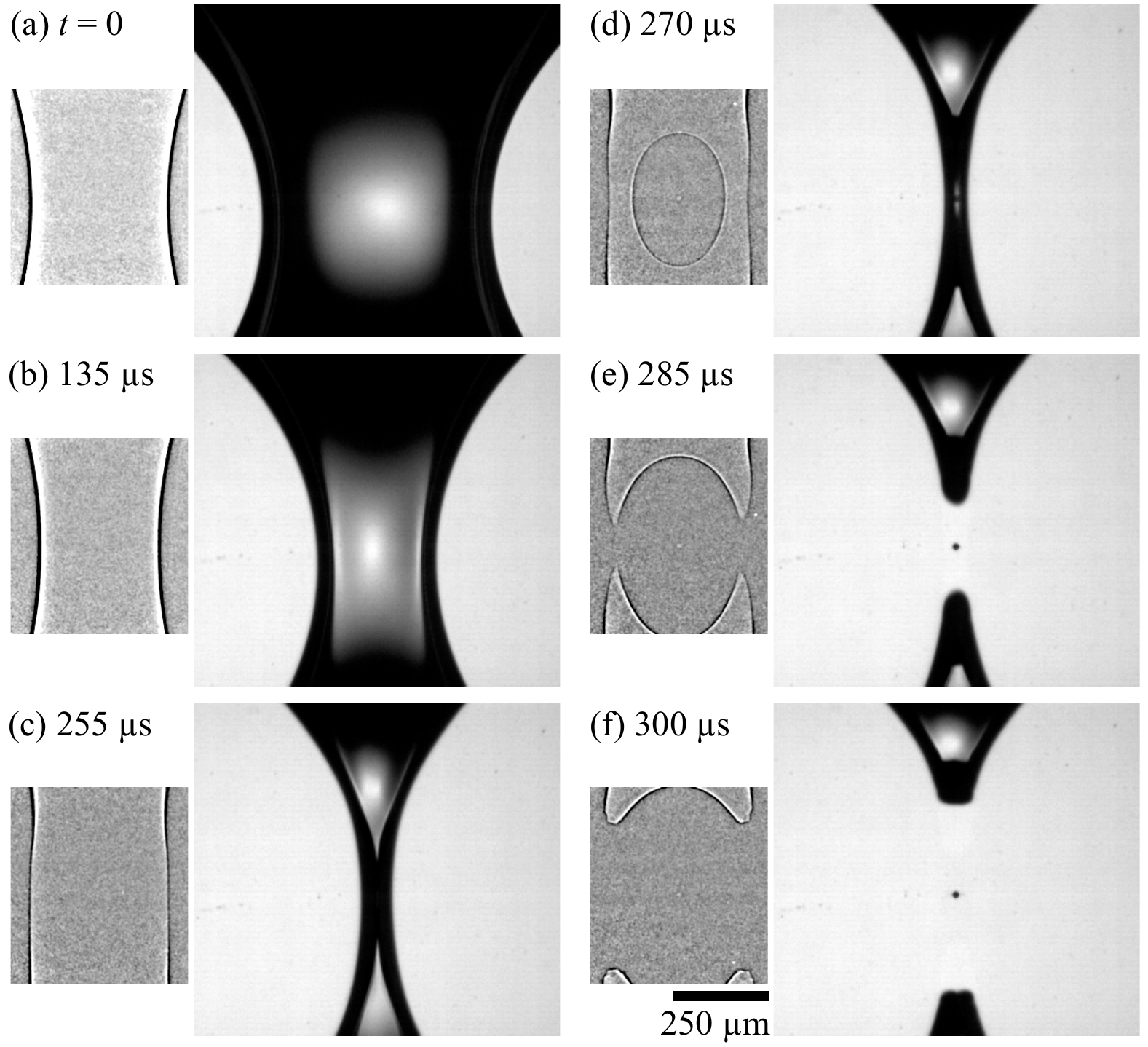}}
\caption{Coalescence-mode of pinch-off observed from orthogonal directions with X-rays (left image of each pair) and visible light. The azimuthal perturbation is created by a burst of He gas from a slot-shaped nozzle. (a)--(d) show the neck profile in visible light thinning dramatically, while in X-rays it contracts only slightly and then expands. The aspect ratio of the neck cross-section continues to increase until the front and back of the neck make contact, leaving the hole seen in the X-ray image of (d). The process leaves a large satellite bubble, most evident in (e) and (f), implying smooth contact at coalescence that prevents the complete drainage of gas~\citep{keim11}. The scale bar is for all images, and corresponding frames are synchronous within 2~$\mu$s.}
\label{fig:xraycoalescence}
\end{figure}
\begin{figure}
\centerline{\includegraphics[width=13cm]{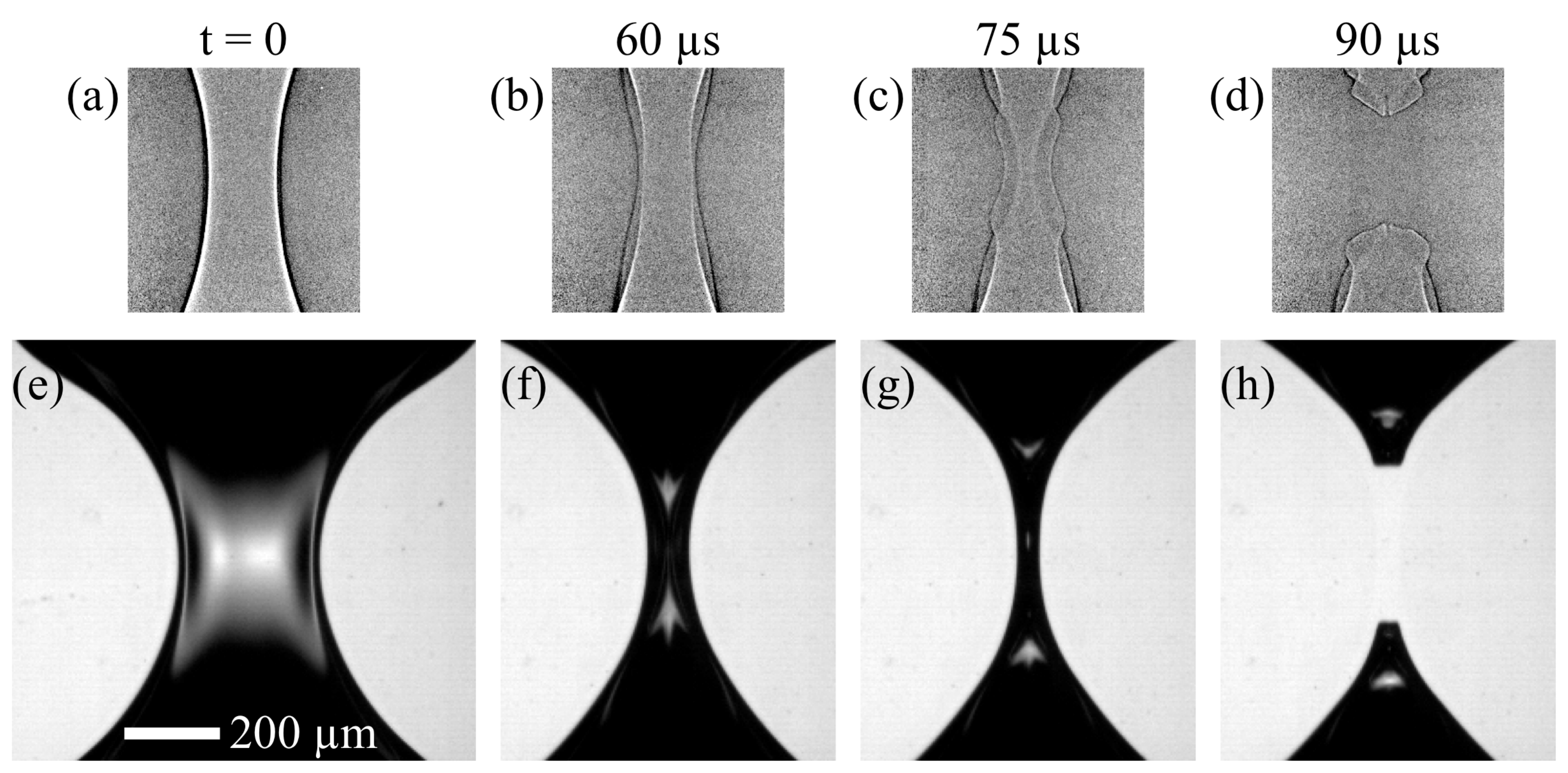}}
\caption{Concavity oscillations observed in experiment. A burst of He gas from a slot-shaped nozzle creates a moderate sized perturbation. (a)-(d): (images captured with X-rays) neck profiles shown from the side. (e)-(h): (images captured with visible light) neck profiles shown from the side in a direction orthogonal to that of (a)-(d). The scale bar shows the length scale for all images. Two images in the same column are close in time but not synchronized due to the experimental limitation. As the neck size shrinks to the same order as the perturbation, the initially convex neck shape, which also obtains in cylindrically-symmetric pinch-off, develops concavities. The complex X-ray image at $t=75$ $\mu s$ shows that the horizontal cross-sections at the top/bottom and middle parts of the neck have strong concavities (observed as four vertical fringes --- two on each side). However, between the top/bottom and middle parts, where the two fringes on each side cross, the cross-sections only have a weak concavity (or no concavity). Our simulation suggests that such an oscillation (strong concavity $\rightarrow$ weak/no concavity $\rightarrow$ strong concavity) is dictated by the contribution from higher modes ($n>2$).}
\label{fig:xraymemory}
\end{figure}

Figure \ref{fig:xraycoalescence} shows X-ray (1st and 3rd columns) and visible light (2nd and 4th columns) observations (side views from two orthogonal directions) of the coalescence-mode break-up. A burst of He gas from a slot-shaped nozzle creates the azimuthal perturbation to the bubble neck. However, figure \ref{fig:xraymemory} shows X-ray and visible light observations of a type of breakup that is produced by a very similar perturbation, yet is completely different in outcome. The complex outcome, especially as observed in X-ray images, is not apparent in experiments where the bubble is inflated quasi-statically \citep{schmidt09,keim11}; there, perturbations are small compared to the observable range of neck sizes. Here, the larger perturbation created by the burst from the slot-shaped nozzle gives rise to an observably concave cross-section that continues to oscillate. 

Although the $n=2$ Fourier mode is the major contribution to the initial perturbation, other higher modes $n=4, 6, 8, ...$ can be generated through nonlinear interactions. In the context of the shape vibrations, the fact that cross-sections at the center and the ends of the neck create four fringes (two on the left side and two on the right side) in X-ray images requires a peculiar concave cross-section (e.g., the third image in figure \ref{fig:2dmodel} (b)). However, as observed in the frame at $75$ $\mu s$ in figure \ref{fig:xraymemory}, moving from top to bottom, the two fringes on each side cross each other. This crossing happens at two heights in the image (c) of figure \ref{fig:xraymemory}, one just above the minimum and one just below the minimum. From top to bottom, the continuity of each of the four fringes indicates that such a crossing cannot be simply explained by a dynamics dominated by a single $n=2$ mode and the reversal of cross-section orientations predicted by Turitsyn~\emph{et al.} \citep{turitsyn09}. An alternative explanation takes into account the effects of higher Fourier modes. The basic idea is that, as indicated by equation (\ref{linstab_phase}), higher modes oscillate with higher frequencies. Thus before an orientation change which is mainly controlled by $n=2$ mode, the interface can go through one cycle of oscillation due to the effects of higher Fourier modes and thus produce different patterns in the phase-contrast image. A comparison between our simulation results and experiments based on this idea will be discussed in Section \ref{subsec:nump2}. In any case, we note that X-ray imaging is uniquely
suited to the geometry seen here: visible-light imaging, even viewing the cavity from above, would be unable to capture the distinct cross-sections from top to bottom in the single $t=75$~$\mu$s frame of figure \ref{fig:xraymemory}, since the minimum $R(\theta)$ of the outer cross-section would coincide with and obscure the maximum $R(\theta)$ of the inner shape.

\subsection{X-ray observations of the tilted jet}

\begin{figure}
\centerline{\includegraphics[width=3in]{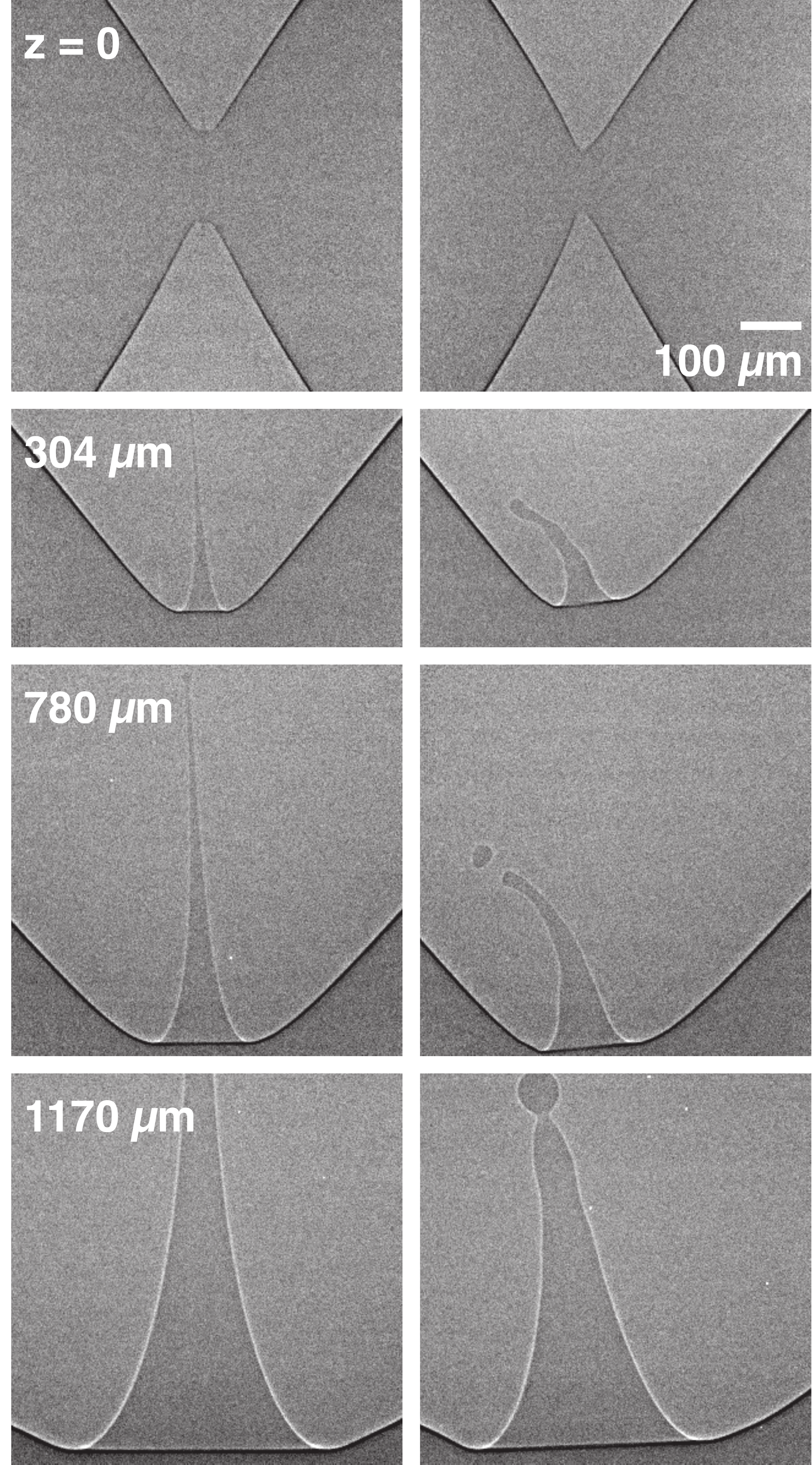}}
\caption{X-ray phase-contrast images of the water jet produced after quasi-static bubble pinch-off from a 6~mm-diameter nozzle. Pinch-off from a level nozzle is shown in the left column, and pinch-off from a nozzle tilted by 0.75$^\circ$ counter-clockwise is on the right. From top to bottom, the images form a sequence in time; however, in order to follow the progression of the upper jet, images are from separate movies taken at increasing heights above the location of pinch-off. From top to bottom, the images are taken at heights at pinch-off $z=0$, then $304$ $\mu$m above pinch-off, then also $780$ $\mu$m and $1170$ $\mu$m above.  Level and tilted images were chosen to correspond as closely as possible. The phase-contrast images show that the small tilt dramatically changes the early evolution of the jet, resulting in a blunted, irregular, and curved jet.  At greater tilts, the water jet collides with the inside of the bubble. The jet's sensitivity to tilts is such that it is not entirely reproducible between movies, presumably due to external vibrations --- as evidenced by the slightly canted jet in the intermediate left-hand images.}
\label{fig:xrayjet}
\end{figure}

While visible-light images show how tilting breaks the symmetry of the overall
neck shape \citep{keim06, keim11}, X-ray images of tilted pinch-off reveal a much more dramatic
response to the tilting perturbation: disruption of the internal Worthington
jet. This jet is due to a strong focusing of liquid flow as the cavity fills in
above and below the pinch-off point \citep{gekle09} (please also see \citep{eggers08} for a review on liquid jets). Our experiments show that
this phenomenon is strongly affected by seemingly unremarkable inclinations of
the nozzle. Figure \ref{fig:xrayjet} shows how a tilt of 0.75$^\circ$ disrupts
the early stages of jet formation; in later stages, the jet still has a slight
tilt. We also observe that this broken symmetry causes the jet's profile to be
knobby and irregular, which is likely the cause of the earlier breakup of the
tilted jet. Experiments at greater tilts show the jet colliding with the inside
of the bubble at early times. Pinch-offs with shape vibrations caused by azimuthal perturbations do not show a similar disruption of the jet when observed with X-rays. These results are further evidence that the inclination of the nozzle
cannot be neglected in any generalized description of pinch-off.

\section{Problem Formulation and Weakly Nonlinear Analysis}\label{sec:prob}

\subsection{Governing equations, boundary and initial conditions}

We model the break-up of an underwater bubble using an idealized scenario first proposed by Longuet-Higgins \emph{et al.}~\citep{longuet91, oguz93}.  The basic assumption is that the bubble neck varies slowly along its length.  As a result the evolution of the neck reduces to a set of two-dimensional problems at the leading order.  The evolution of the cross-section shape at a particular height is decoupled from the shape evolution at other heights.  Gas flow through the bubble neck, surface tension effects and viscous drag are neglected. Despite these simplifications, this model successfully reproduces the measured neck shape evolution when the dynamics is nearly cylindrically-symmetric.  It also reproduces the measured instability dynamics \citep{schmidt09,keim11}. 

\begin{figure}
\centerline{\includegraphics[width=12cm]{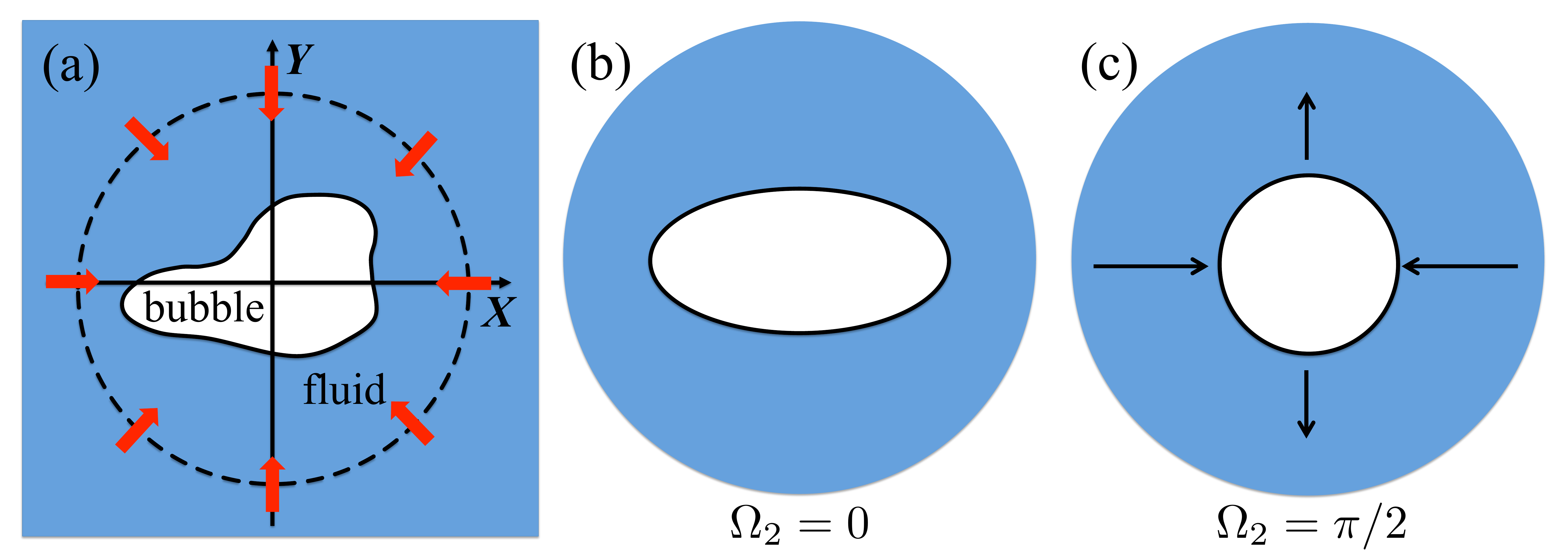}}
\caption{(a) Schematic of two-dimensional model: cross-section of the bubble neck at the minimum contracts inwards due to a radially symmetric influx of water from the far field. Parts (b) and (c) illustrate two different initial conditions.  (b) Azimuthal perturbation in the form of an $n=2$ vibration with phase $\Omega_2 =0$.  At the initial moment, the velocity field is at rest but the bubble cross-section is elongated.  (c) Same $n=2$ perturbation but with phase $\Omega_2 = \pi/2$.  At the initial moment, the velocity field contracts along the $x$-direction while elongates along the $y$-direction.  The cross-section of the bubble is unperturbed and circular.}
\label{fig:setup}
\end{figure}

Within this model, the bubble interior is simply a region of uniform pressure $P(t)$.  The exterior velocity field is irrotational and incompressible.  Equivalently, the velocity field $\bf{u}$ is given by the gradient of a velocity potential $\Phi$ satisfying Laplace's equation
\begin{equation}\label{laplace}
\nabla^2 \Phi =0
\end{equation}
The bubble surface is given by the curve $S(r, \theta, t)$ specifying its cross-section shape at a given height $z$, as illustrated in figure \ref{fig:setup} (a).  Far from the bubble neck cross-section, the velocity field in the exterior is a radial inflow,
\begin{equation}
\label{farflow}
{\bf{u}} = -\frac{Q}{r} {\bf{e}_r}\qquad r \rightarrow \infty
\end{equation}
To keep the analysis simple, we hold the volume flux $2\pi Q$ constant over time.  
This is equivalent to prescribing that the average radius of the cross section $\bar{R}(t)$ decreases as $\bar{R}(0)\sqrt{(t_*-t)/t_*}$, where $t_*=\bar{R}(0)/2Q$ is the time when the average size of the bubble goes to zero. The square-root decrease reproduces, at leading order, the measured $(t_*-t)^{0.56}$ decrease in $\rmin$ \citep{burton05,keim06}. 

The kinematic boundary condition states that a point on the bubble surface moves with the fluid surface:
\begin{equation}{\label{eqn:kinematicbc}}
\frac{d\bf{x}}{dt}= \left(\frac{\partial}{\partial t}+\nabla\Phi\cdot\nabla \right){\bf{x}}=\nabla \Phi|_S
\end{equation}
The normal stress condition at the bubble surface $S({\bf{x}}, t)$ has the form 
\begin{equation}{\label{eqn:dynamicbc}}
\rho \left [\frac{\partial \Phi}{\partial t}+\frac{1}{2}|\nabla\Phi|^2 \right ]_S = -P(t)
\end{equation}
Given the bubble cross-section shape and the velocity potential on the surface at the initial moment,  equations (\ref{eqn:kinematicbc}) and (\ref{eqn:dynamicbc}) describe how they evolve over time.  

We examine a particularly simple set of initial perturbations.  At $t=0$, the bubble cross-section shape $r = S(\theta, t)$ has the form\begin{equation} \label{icshape}
S(\theta, t=0) = R_0 + A_2 \cos \Omega_2 \cos(2 \theta)
\end{equation}
The initial state corresponds to a circular cross-section with initial radius $R_0$ distorted by a single $n=2$ vibrational mode with amplitude $A_2$ and phase $\Omega_2$.  Turitsyn, Lai and Zhang \citep{turitsyn09} held $\Omega_2$, the phase of the Fourier mode at the initial moment, constant and monitored how the break-up dynamics changes as $A_2$, the amplitude of the perturbation, is varied.  Here we keep $A_2$ constant and vary $\Omega_2$, as illustrated in figure \ref{fig:setup} (b) and (c).  This choice allows us to explore how the final break-up dynamics varies as a function of initial conditions while maintaining approximately the same dynamic range in $\rmin$ for the different initial conditions.  We also use a boundary integral simulation, described in more detail in Section \ref{sec:num}, instead of the spectral method used by Turitsyn, Lai and Zhang \citep{turitsyn09}.  This approach is capable of tracking the interface evolution even when the cross-section becomes re-entrant. 

In the following sections, unless written out explicitly, all the quantities are understood as being rescaled by the initial length scale $R_0$, time scale $R_0^2/Q$, or their combinations.

\subsection{Reformulation in canonical variables}\label{subsec:RV}

The pinch-off dynamics, or equations (\ref{laplace}) to (\ref{eqn:dynamicbc}), corresponds to a Hamiltonian evolution. Here we follow an approach developed by Dyachenko \emph{et al}. in the context of wave breaking \citep{dyachenko96,zakharov02} in which the governing equations are rewritten in terms of canonical variables that display the Hamiltonian structure clearly.  To do so, we first transform the bubble surface in real space $S(r, \theta, t)$ onto a unit circle on the complex plane $w$.  A point on $w$ plane is related to the location ($r$, $\theta$) on the 2D plane of the cross section by $z(w,t)=re^{i\theta}$. The exterior of the bubble is mapped into the exterior of a unit circle. Next, instead of solving directly for the evolution of the interface and the velocity potential, we use the variable transformation 
\begin{equation}
{\cal R}(w, t)=1/(w\partial_wz)\quad\quad{\cal V}(w, t) =\partial_w\Psi/\partial_wz
\label{defRV}
\end{equation}
where $\Psi(w, t)$ is the complex velocity potential governing the exterior flow.  The variable ${\cal V}$ corresponds to the speed of the fluid on the interface. The variable $\cal R$ has no straightforward physical interpretation, though it is clearly related to how distorted the bubble cross-section shape is relative to the unit circle on the $w$-plane. In terms of these new variables, the evolution equations for the interface (equation (\ref{eqn:kinematicbc})) and for the velocity potential (equation (\ref{eqn:dynamicbc})) now have the form
\begin{equation}
\partial_t{\cal R}=w(\partial_w{\cal R}){\cal A}\{Re[{\cal R}{\cal V^*}]\}-w{\cal R}\partial_w{\cal A}\{Re[{\cal R}{\cal V^*}]\}
\label{Reqn}
\end{equation}
\begin{equation}
\partial_t{\cal V}=w(\partial_w{\cal V}){\cal A}\{Re[{\cal R}{\cal V^*}]\}-w{\cal R}\partial_w{\cal A}\{\frac{|{\cal V}|^2}{2}\}
\label{Veqn}
\end{equation}
where the $*$ symbol denotes the complex conjugation. The operator ${\cal A}$ is the Cauchy integral
\begin{equation}
{\cal A}\{f\}(w)=\frac{1}{2\pi i}\oint_{|w'|=1}\frac{dw'}{w'}\frac{w+w'}{w-w'}f(w')
\label{Aintegral}
\end{equation}
This operation takes a  real-valued function $f(\omega)$ that is defined along the unit circle $\omega = e^{i \alpha}$ and returns a complex function ${\cal A}{f}(\omega)$ which is analytic everywhere in the exterior of the unit circle \citep{carrier,shraiman84}. 

\subsection{Weakly Nonlinear Analysis}\label{subsec:weaklynonlinear}

Turitsyn, Lai and Zhang \citep{turitsyn09} showed that the cylindrically-symmetric pinch-off is pre-empted by a coalescence-mode break-up and can be recapitulated by simply assuming that interface evolution preserves the linear stability dynamics down to the moment of topology change. With this assumption, the perturbation mode amplitude remains constant in time while its phase winds up faster and faster as pinch-off approaches. Specifically, in ${\cal R}$, ${\cal V}$ variables, this assumption means that, a perturbation in the form of an $n=2$ Fourier mode with a small initial amplitude evolves as 
\begin{eqnarray}\label{n2only}
{\cal R} &=& \frac{1}{\omega} \left (c_0(t) +  \frac{c_2(t)}{\omega^2} \right) \\
{\cal V} &=& \frac{1}{\omega} \left (d_0(t) +  \frac{d_2(t)}{\omega^2} \right)
\label{n2only2}
\end{eqnarray}
The coefficients are related to the coefficients in the standard Fourier mode expansion for the bubble surface and the velocity field as follows: $c_0(t) = 1/ \bar{R}(t)$ where $\bar{R}$ is the average radius of the bubble neck cross-section, and the coefficient $d_0 (t) = -1 / \bar{R}(t)$.  The $n=2$ mode coefficients are given by\begin{eqnarray}\label{eqn:c2}
c_2 (t) &=& A_2 \cos \phi_2(t)/ {\bar{R}}^2(t) \\
d_2(t) &=& - A_2 \sin \phi_2(t)/ \bar{R}^2(t)
\label{c2d2}
\end{eqnarray}
The linear stability behavior, in which the relative amplitude of the $n=2$ mode becomes more pronounced as the break-up proceeds (equation (\ref{linstab_amp})), is reflected in the $\bar{R}^{-2}(t)$ divergences of $c_2$ and $d_2$ (equations (\ref{eqn:c2}) and (\ref{c2d2})). Formulae (\ref{n2only}) to (\ref{c2d2}) predict evolution into a coalescence-like pinch-off.  Moreover, they approximately reproduce  the trend from the numerical simulations:  changing the initial amplitude of the perturbation changes when the coalescence happens, and therefore the orientation of the coalescence plane.   

While successful as a first approximation, there are systematic discrepancies between predictions based on linear stability analysis (equations (\ref{n2only}) and (\ref{n2only2})) and the simulation results.  The most obvious one is the shape of the bubble cross-section at contact.  In the simulation, a variety of final shapes are seen for coalescence-mode break-up.  This variety reflects the fact that higher harmonic modes, which are not included in the linear evolution model, are generated via nonlinear resonance as break-up approaches its final stages. We therefore perform a weakly nonlinear analysis where we track the evolution of both the initial perturbation mode, which has the form of an $n=2$ Fourier mode, and the first higher harmonic, which is the $n=4$ Fourier mode.  We do this not only because it is a natural extension of the linear stability analysis but also because the nonlinearity associated with the interface evolution has the property of generating only higher harmonics from its initial perturbation mode but not subharmonics.  This feature is most evident from the form of the evolution equations in ${\cal R}$, ${\cal V}$ variables (equations (\ref{Reqn}) and (\ref{Veqn})).  Thus the evolution with a single $n=2$ mode initial perturbation should give a general idea of how the surface evolution proceeds for all perturbations comprised of a single Fourier mode, whatever the value of $n$. 

Assuming that $A_2 \ll \bar{R}$, $A_4 \propto A_2^2$, and neglecting all terms that are higher order than $A_2^2$, the full evolution equations are simplified to the following equations for the amplitudes $A_0$, $A_2$, $A_4$, and phases $\phi_2$ and $\phi_4$
\begin{eqnarray}
\frac{\rd A_0}{\rd t}&=&-\frac{1}{A_0}-\frac{A_2^2}{A_0^3}\cos(\phi_2)\sin(\phi_2)\label{Aeqn1}\\
\frac{\rd A_2}{\rd t}&=&0\\
\frac{\rd \phi_2}{\rd t}&=&\frac{1}{A_0^2}\\
\frac{\rd A_4}{\rd t}&=&\frac{A_2^2}{3A_0^3}[(\cos^2(\phi_2)+\cos(\phi_2)\sin(\phi_2))\cos(\phi_4)\nonumber\\&&+(3\sqrt{3}\sin^2(\phi_2)+\sqrt{3}\cos(\phi_2)\sin(\phi_2))\sin(\phi_4)]\\
\frac{\rd \phi_4}{\rd t}&=&-\frac{A_2^2}{3A_4A_0^3}[(\cos^2(\phi_2)+\cos(\phi_2)\sin(\phi_2))\sin(\phi_4)\nonumber\\&&-(3\sqrt{3}\sin^2(\phi_2)+\sqrt{3}\cos(\phi_2)\sin(\phi_2))\cos(\phi_4)]+\frac{\sqrt{3}}{A_0^2}\label{AeqnL}
\end{eqnarray}
An inspection of the evolution equations for $A_4$ and $\phi_4$ for initial states containing only a single, $n=2$ mode shows that a smooth time evolution for $A_4=0$ at $t=0$ obtains only if $\phi_4$ at $t=0$ satisfies the equation
\begin{equation}\label{eqn:numerator}
[ \cos^2(\phi_2)+\sin(2\phi_2)/2 ]\sin(\phi_4) -
[ 3\sqrt{3}\sin^2(\phi_2)+\sqrt{3}\sin(2\phi_2)/2 ]\cos(\phi_4) = 0
\end{equation}
With this choice, the evolution equation for $\phi_4$ at small $t$ simplifies to
\begin{equation}\label{eqn:initial_phi4}
\frac{\rd \phi_4}{\rd t}=\frac{\sqrt{3}}{A_0^2}
\end{equation}



\begin{figure}
\centerline{\includegraphics[width=13cm]{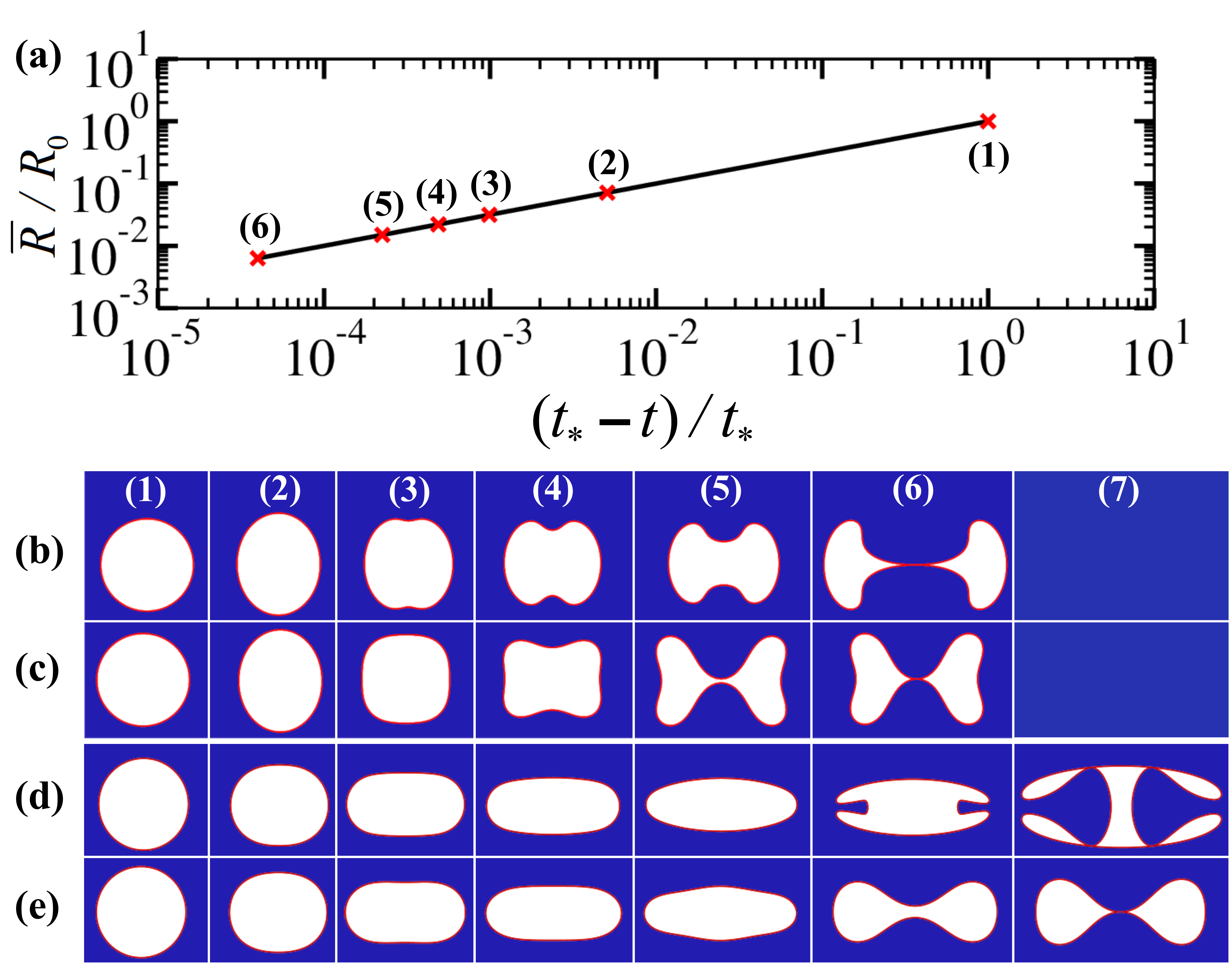}}
\caption{Evolution of the bubble cross-section shape near pinch-off.  (a)  Average radius $\rmin$ as a function of $(t_* - t) /t_*$, where $t_*$ corresponds to the moment of the cylindrically-symmetric pinch-off.  (b) Boundary integral simulation of shape evolution starting with a single $n=2$ mode 
($A_2 = 0.01$ and  $\Omega_2=0.4\pi$).  (c) Weakly nonlinear analysis predictions for initial state $A_2 = 0.01$ and  $\Omega_2=0.4\pi$.  (d) Boundary integral simulation of shape evolution starting with a single $n=2$ mode ($A_2 = 0.01$ and  $\Omega_2=0.9\pi$).  (c) Weakly nonlinear analysis predictions for initial state $A_2 = 0.01$ and  $\Omega_2=0.9\pi$.}
\label{fig:mockWkly}
\end{figure}

Figure \ref{fig:mockWkly} shows snapshots from two representative sets of cross-section shape evolutions. Both sets are calculated using the evolution equations described above but use different values of $\Omega_2$, the initial phase of $n=2$ mode.  As a comparison, we also display results from our boundary integral simulation (Section \ref{sec:num}), whose results do not require higher harmonic mode amplitudes to be small.  The snapshots in figure \ref{fig:mockWkly} (b) and (c) correspond to shape evolutions starting from $\Omega_2 = 0.4 \pi$.  Here the predictions from weakly nonlinear analysis (figure \ref{fig:mockWkly} (b)) agree qualitatively with the boundary integral results (figure \ref{fig:mockWkly} (c)).  In contrast, when the initial phase of the $n=2$ mode has the value $\Omega_2 = 0.9 \pi$, the shape evolutions predicted by the weakly nonlinear analysis initially agree with the simulation, but then deviate dramatically as pinch-off approaches. This trend holds when we compare analysis and simulation results for different $A_2$ values, as well as for single-mode initial states with different mode numbers. 

\begin{figure}
\centerline{\includegraphics[width=13cm]{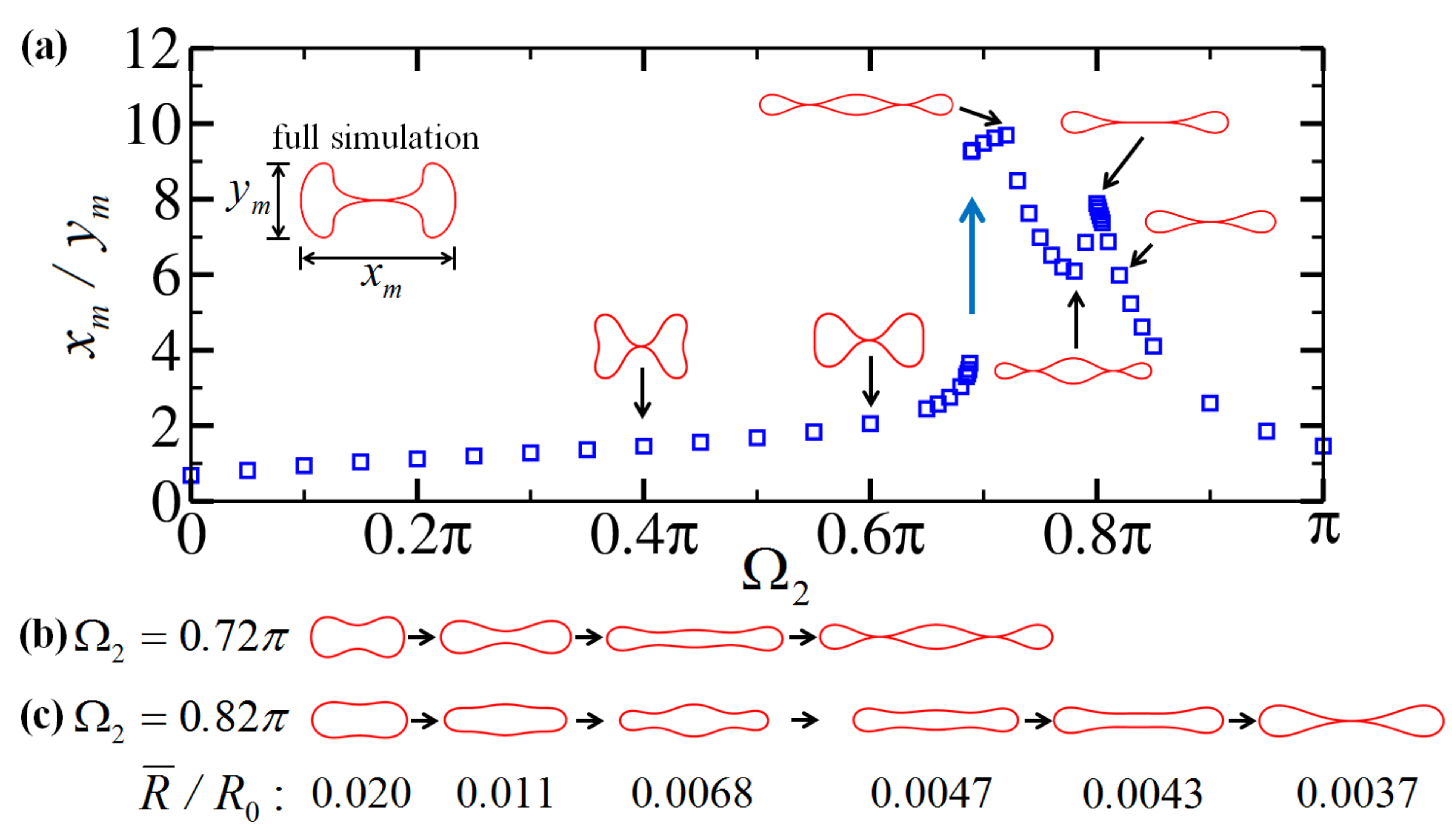}} 
\caption{Weakly nonlinear analysis:  Aspect ratio $x_m/y_m$ of the bubble cross-section at the moment of contact as a function of $\Omega_2$, the initial phase of the $n=2$ Fourier mode perturbation. The initial perturbation amplitude is fixed at $A_2=0.01$. The aspect ratio $x_m/y_m$ first rises smoothly towards a peak value as $\Omega_2$ increases, then abruptly jumps to around its peak value at about $\Omega_2=0.7\pi$. As $\Omega_2$ keeps increasing, the aspect ratio first decreases, and then increases before it decreases again, forming a secondary peak at around $0.8\pi$. Rows (b) and (c) show that this rapid non-monotonic variation is caused by the different interference patterns between modes $n=2$ and $n=4$. (b) Around $\Omega_2=0.72\pi$, the destructive interference between modes $2$ and $4$ gives rise to a peak in the aspect ratio $x_m/y_m$ and transforms the final outcome from the single-point contact to the double-point contact. (c) Increasing the value of the initial phase $\Omega_2$ further gives rise to a shape evolution which just misses the double-point contact and finally ends up with a single-point contact.}
\label{fig:wklynonlinear} 
\end{figure}

In order to gain insight into why the evolution remains dominated by the single, initial mode (and are therefore well approximated by predictions from weakly nonlinear analysis) for only some initial states, we need to quantify the discrepancy between the simulations and the weakly nonlinear analysis as a function of  $\Omega_2$.  Figure \ref{fig:wklynonlinear}(a) plots the final value of the bubble cross-section aspect ratio.  This is defined as a ratio of the maximum lateral extent $x_m$ divided by the maximum extent $y_m$ along the orthogonal in-plane direction.  Because the $n=2$ perturbation is both left-right and up-down symmetric and the shape evolution preserves spatial symmetry, we show only results from the interval $0 \leq \Omega_2 \leq \pi$. The range $ \pi < \Omega_2 < 2 \pi$ produces an identical set of outcomes,  except that the coalescence-plane orientation is rotated by $90^\circ$.  

When modes $2$ and $4$ are in constructive interference, meaning that mode $4$ promotes the coalescence-mode break-up by pushing the top, bottom, left and right sides of the interface inwards, the final aspect ratio shows a weak variation with $\Omega_2$. The exception occurs in the interval between around $0.7\pi$ and $0.9\pi$. These are also initial states for which we find the greatest discrepancies between the weakly nonlinear analysis and full boundary integral simulations.  

The unusual double peak structure in the $x_m/y_m$ versus $\Omega_2$ plot owes its existence to precise interactions between the cylindrically symmetric implosion and the shape vibrations. Specifically, when the initial phase first approaches $0.7\pi$, the aspect ratio first starts to increase rapidly, followed by an abrupt jump as the final cross-section shape changes from one-point contact to a double contact. The second peak around $0.8\pi$ forms when the contact type switches from two-point contact back to one-point contact.  During this process $y_m$ switches from being dominated by the central lobe to being dominated by the side lobes. This change in behavior occurs because the $n=4$ shape oscillation is now in destructive interference with the $n=2$ oscillation.  In other words, the motion of the $n=4$ mode now delays topology change by holding the top and the bottom surfaces away from each other.  The net effect is to cause the final aspect ratio $x_m/y_m$ to increase and to create a double contact (figure \ref{fig:wklynonlinear} (b)).  

Increasing $\Omega_2$ above $0.8 \pi$ introduces another change in the break-up dynamics (figure \ref{fig:wklynonlinear} (c)).  For these values, the $n=4$ shape oscillation completes an orientation reversal as the pinch-off approaches.  This means the $n=2$ and $n=4$ modes switch from destructive to constructive interference.  As a consequence, the cross-section evolves close to a double contact but does not complete it. Instead the cross-section broadens out into an oblong shape.  The oblong structure first forms a central lobe, then narrows inwards at the midpoint and finally pinches off via a single-point coalescence. 

To recap, weakly nonlinear analysis tells us that the evolutions of initial states remain simply dominated by the single mode initially present if the higher harmonic mode generated by nonlinear interaction remains in constructive interference with the original mode.  Otherwise, simulations consistently show more complicated shape evolutions.  This suggests that the generation of the higher harmonic modes proceed differently when the modes are in destructive interference.  To test this idea, we turn to boundary integral simulations capable of resolving the generation and the evolution of many high harmonic modes from the single mode originally present. 

\section{Numerics}\label{sec:num}

\subsection{Reformulation of Governing Equations as Boundary Integral}

While reformulating the governing equations in terms of $\cal R$ and $\cal V$ allows expedient numerical solution via series expansion \citep{turitsyn09}, it also carries a major disadvantage. The numerical method based on the series expansion fails to resolve re-entrant regions along the interface. But a subset of break-up dynamics involves the formation of a re-entrant cross-section shape. To investigate these break-up modes, we turn to a different formulation of the governing equations.  

In our model, the exterior flow is assumed to be irrotational and inviscid, therefore the velocity field is given by the gradient of a potential satisfying Laplace's equation.  The fact that the bulk flow satisfies a linear governing equation allows a Green's function approach \citep{pozrikidis}.  In essence, this represents the normal stress distribution along the bubble surface (\ref{eqn:dynamicbc}) as a distribution of force monopoles (Stokeslets), and dipoles whose strengths depend on the velocity distribution on the surface.  They in turn determine how the velocity field everywhere evolves. At the bubble surface $S$, this boundary integral approach yields an integral equation relating the velocity potential $\Phi$ at a point $\bf{x_0}$ to the potential $\Phi(\bf{x})$ and its derivative everywhere along the surface, 
\begin{equation}
\frac{1}{2}\Phi({\bf x_0})=-\oint_{{\bf x}\in S}G({\bf x_0}, {\bf x})({\bf n}\cdot{\bf \nabla})\Phi({\bf x})ds({\bf x}) + \oint_{{\bf x}\in S}\Phi({\bf x})({\bf n}\cdot{\bf \nabla})G({\bf x_0}, {\bf x})ds({\bf x})
\label{integraleqn}
\end{equation}
where $G({\bf x_0}, {\bf x})$ is the free-space Green's function solution to the Laplace's equation in 2D and has the form
\begin{equation}
G({\bf x_0}, {\bf x})=-\ln(|{\bf x_0}-{\bf x}|)/2\pi
\label{greensfunction}
\end{equation}
The symbol ${\bf n}$ in equation (\ref{integraleqn}) is the surface normal at location ${\bf x}$, defined as conventional in this context with the positive value corresponding to the direction into the exterior.  

With this reformulation, the time-evolution of the bubble cross-section requires only a quadrature over the bubble surface instead of solving for the velocity potential everywhere in the exterior.  It therefore provides a significant saving in computation cost.  More importantly this approach allows us to examine re-entrant cross-section shapes. Equation (\ref{integraleqn}) together with the kinematic boundary condition (\ref{eqn:kinematicbc}) and the dynamic boundary condition (\ref{eqn:dynamicbc}) that describes the time-evolution of the velocity potential $\Phi(t)$ fully specify the time-evolution of our system. 

\subsection{Numerical Methods}

We implement the boundary integral method numerically by discretizing the interface via an adaptive mesh.   After experimenting with redistribution schemes that move grid points according to curvature or arc distance, we found that the most stable scheme is also the simplest:  we begin with the grid points evenly distributed along the surface, then allow the points to move with the surface velocity. This scheme concentrates points in regions where the surface is highly curved, and maintains a good spatial resolution when those regions become re-entrant afterwards.  

The evolution dynamics preserves the spatial symmetry imposed by the initial Fourier mode perturbation.  A bubble neck cross-section perturbed by a single $n=2$ Fourier mode remains both left-right and up-down symmetric over time. When solving the time-evolution of the cross-section, we enforce this symmetry to the cross-section to stabilize and accelerate our simulation. A typical run uses $800$ grid points. We have checked that increasing the total number of points, or decreasing it by a factor of $2$, does not change the results reported here. 

We use an explicit 4th order Runge-Kutta time-stepping scheme with varying time-step size to update the interface. The time-step size $\Delta t$ is prescribed as $C|\Delta x_{min}|^2$ where $C$ (typically $200$ for $800$ grid points) is chosen empirically and $|\Delta x_{min}|$ is the minimum spacing between adjacent grid points. Increasing or decreasing the value of $C$ doesn't change the results reported here. For some initial conditions, when the cross-section is close to a topology change or when the surface curvature changes rapidly, we reduce $C$ to $100$ or $50$ during our simulation to obtain more data in respective time windows.

The simulation halts either when the cross-section shape completes a topology change, or when the surface becomes too highly curved, specifically when the curvature in the neighborhood of a grid point is comparable to the grid spacing.  More details about the numerical implementation can be found in appendix B. 

\subsection{Simulation Results:  $n=2$ perturbations}\label{subsec:simulationresults}

\begin{figure}
\centerline{\includegraphics[width=13.5cm]{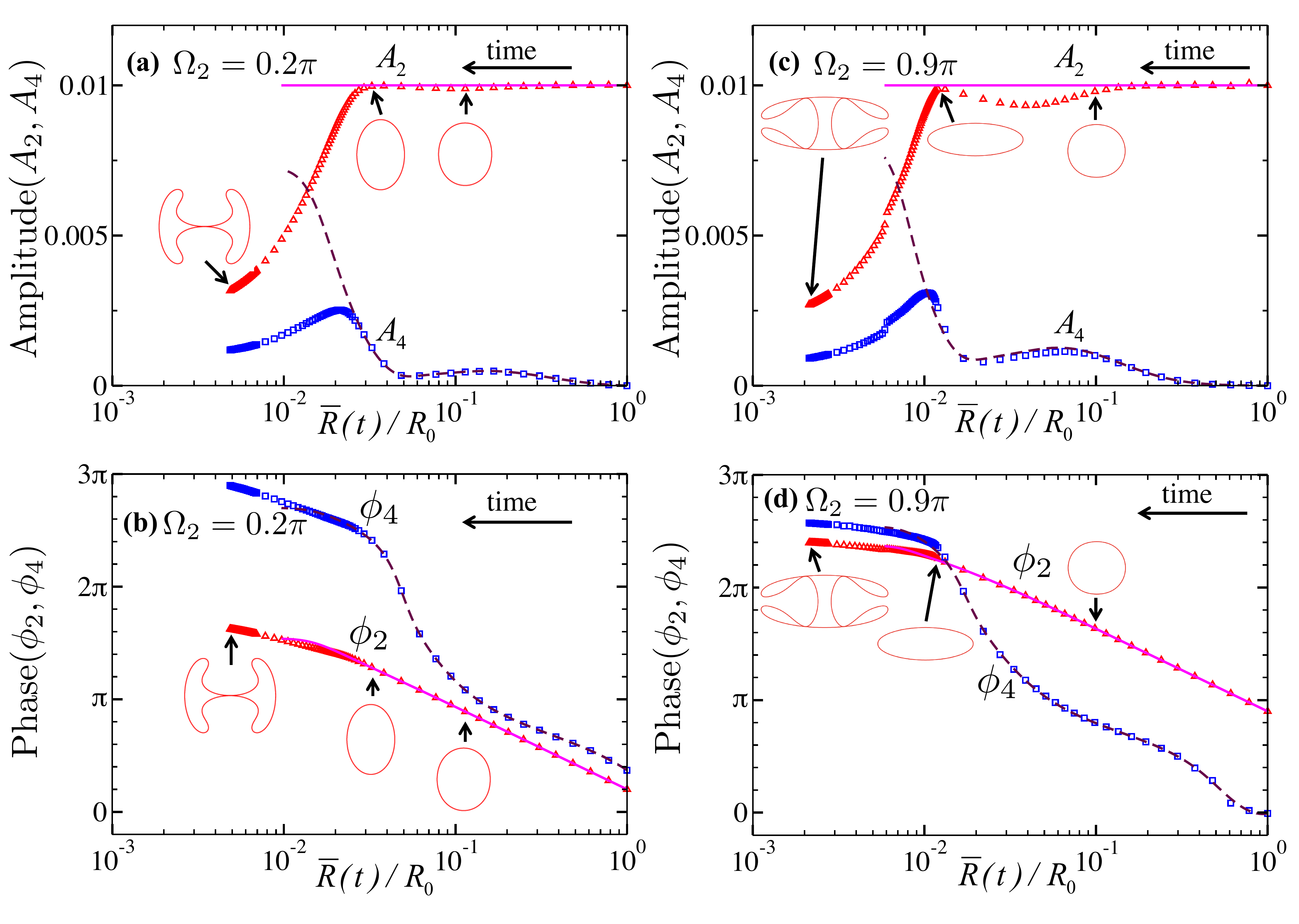}}
\caption{Evolution of the initial $n=2$ perturbation together with the $n=4$ mode, the first harmonic mode generated by nonlinear interactions. The symbols (color online) are results from the boundary integral simulation.  The lines are predictions from the weakly nonlinear analyis.  (a) Time evolution of mode amplitudes $A_2$ and $A_4$ for an initial state containing a single $n=2$ mode with amplitude $A_2 = 0.01$ and phase $\Omega_2 = 0.2 \pi$.  (b) Time evolution of  phase variables $\phi_2$ and $\phi_4$ for the same initial state. 
(c) Time evolution of mode amplitudes $A_2$ and $A_4$ for an initial state containing a single $n=2$ mode with amplitude $A_2 = 0.01$ and phase $\Omega_2 = 0.9 \pi$.  (d) Time evolution of  phase variables $\phi_2$ and $\phi_4$ for the same initial state.
}
\label{fig:mode2and4compare}
\end{figure}

Two shape evolutions calculated using the boundary integral code are given in figure \ref{fig:mockWkly}.  As discussed earlier, a comparison between the simulation results and predictions from the weakly nonlinear analysis shows that the two methods produce qualitatively different results for some initial conditions. Figure \ref{fig:mode2and4compare} displays the time evolution of the mode amplitudes and phases for the same two initial states.  To make this comparison, we fit the output of the boundary integral simulation, i.e., the velocity potential $\phi(t)$ and the surface shape $S(t)$ to a truncated expansion in $\cal R$, $\cal V$ variables
\begin{eqnarray}
{\cal R} &=& \frac{1}{\omega} \left (c_0(t) +  
\Sigma_{n=2}^{N} \frac{c_n(t)}{\omega^n} \right) \\
{\cal V} &=& \frac{1}{\omega} \left (d_0(t) +  
\Sigma_{n=2}^{N} \frac{d_n(t)}{\omega^n} \right) . 
\end{eqnarray}  
Typically we keep $20$ terms in the expansion.  Using more or fewer terms does not change the results reported here. 

Overall, we see good agreement between the weakly nonlinear analysis and the boundary integral simulation for both initial states.  Initially, the amplitude of the perturbation mode is small relative to the size of the cross-section and the evolution of the $n=2$ mode displays the trend we expect from linear stability \citep{schmidt09}.  Its amplitude $A_2$ remains essentially constant while the phase $\phi_2$ winds up as a linear function of $\ln(R_0/\bar{R})$.  Later, $A_2$ decreases.  In contrast, $A_4$ grows to a maximum value before starting to decrease as well.  Because energy from the break-up is confined to $n=2$ and $n=4$ modes in the weakly nonlinear analysis, the predicted peak in $A_4$ is larger than that found from the simulation, where the break-up energy can be distributed into higher harmonics.  The simulation also ends at a smaller value of $\bar{R}$, signaling that the effect of the higher harmonics also delays the moment of coalescence.  The phase evolutions from the simulation agree closely with those predicted by the weakly nonlinear analysis. More importantly, the boundary integral results agree quantitatively with predictions from weakly nonlinear analysis for both sets of initial conditions.  This demonstrates that the qualitative discrepancy in the shape evolutions we see for $\Omega_2 = 0.9\pi$ owes its existence to the presence of high harmonic modes not included in the weakly nonlinear analysis.

\begin{figure}
\centerline{\includegraphics[width=12cm]{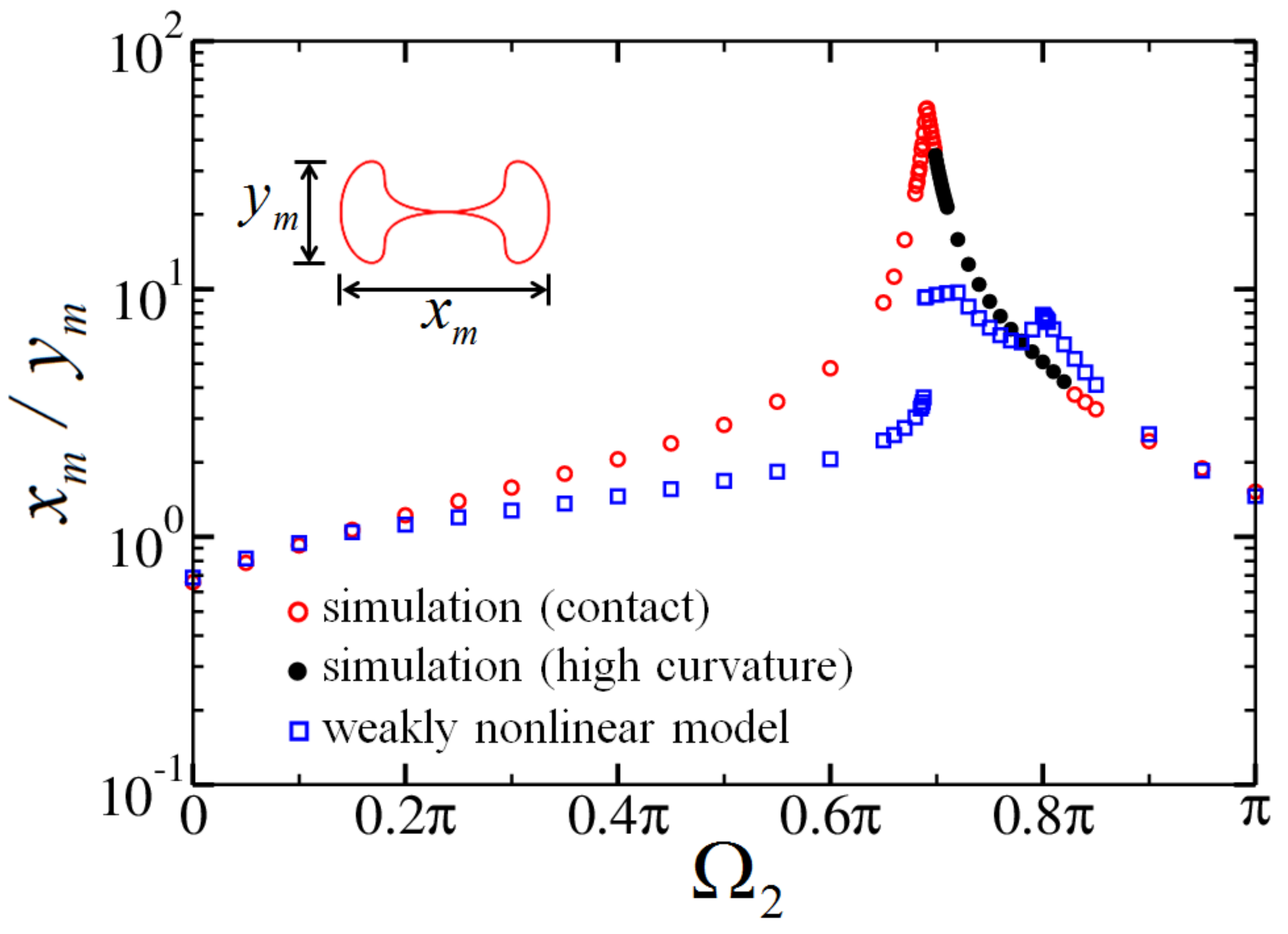}}
\caption{Aspect ratio of the bubble cross-section at pinch-off as a function of the initial phase $\Omega_2$.  The initial perturbation is a single $n=2$ Fourier mode with initial amplitude $A_2 = 0.01$. The black solid circles (color online) correspond to shape evolutions that give rise to highly curved surfaces instead of a coalescence-mode break-up (see figures \ref{fig:seq} and \ref{fig:slot} for examples).}
\label{fig:aspectratioCompare}
\end{figure}

We next examine the second prediction from the weakly nonlinear analysis: the rapid increases and decreases in the final aspect ratio $x_m/y_m$ of the bubble cross-section around the onset of destructive interference between mode $2$ and $4$ (figure \ref{fig:wklynonlinear}).  Figure \ref{fig:aspectratioCompare} plots the aspect ratio of the bubble cross-section from the full simulation versus the initial phase $\Omega_2$.  The trends from the two calculations are qualitatively similar.  Both the boundary integral simulation and the weakly nonlinear analysis show that the pinch-off dynamics produces final shapes that vary rapidly within a relatively narrow range of initial phase values.  However, there are significant quantitative differences.  The double-peak structure predicted by weakly nonlinear analysis is replaced by a single, higher and broader peak in the full simulation.  Moreover we see a qualitatively different set of final outcomes.  These, indicated by black dots in figure \ref{fig:aspectratioCompare}, correspond to final cross-section shapes characterized by the bubble interface developing nearly singular shapes as break-up approaches.  

\begin{figure}
\centerline{\includegraphics[width=13cm]{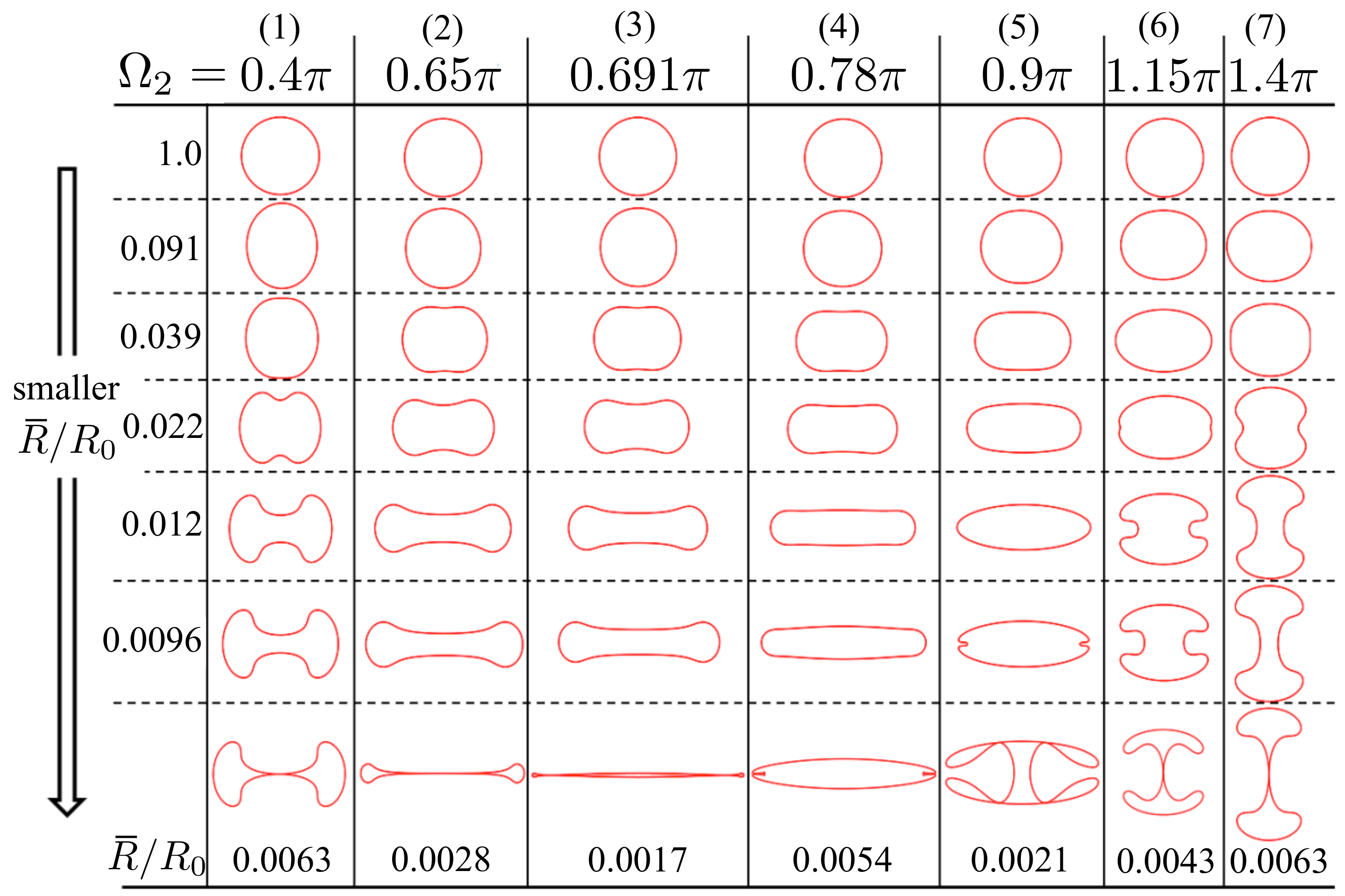}} 
\caption{Changing $\Omega_2$, the initial value of the phase for the $n=2$ vibration, creates a variety of bubble pinch-off dynamics. In order to display clearly how the distortion develops over time, we have rescaled the bubble cross-section shapes by the corresponding values of the average radius $\bar{R}$. The last row plots the cross-section shape at the time when our simulation halts, either due to a topology change, or due to the development of a sharply curved region on the interface. The final values of $\bar{R}/R_0$ are labeled below corresponding cross-section shapes in the last row.}
\label{fig:seq}
\end{figure}

\begin{figure}
\centerline{\includegraphics[width=13cm]{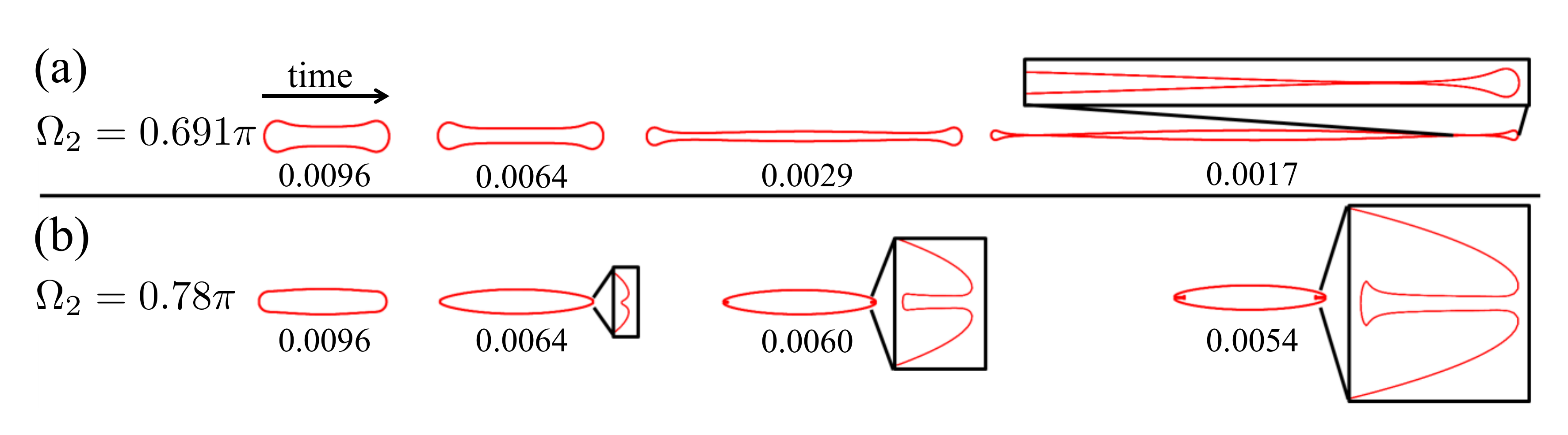}}
\caption{Close-up of surface evolutions resulting in large cross-section aspect ratios.  (a) Double-contact break-up ($\Omega_2 = 0.691 \pi$):  the cross-section of the bubble evolves into a thin slit-like shape with two rounded ends, which then contacts at two distinct points. (b) Formation of water fingers ($\Omega_2 = 0.78 \pi$):  the cross-section of the bubble develops re-entrant regions which then propagate into the air region as straight fingers of water with rounded tips.  Over time, the tip of the finger flattens and two sharp corners develop.  The extremely high curvature at the two corners halts our simulation. The value of $\bar{R}/R_0$ is shown below each corresponding cross-section shape.}
\label{fig:slot}
\end{figure}

\begin{figure}
\centerline{\includegraphics[width=13cm]{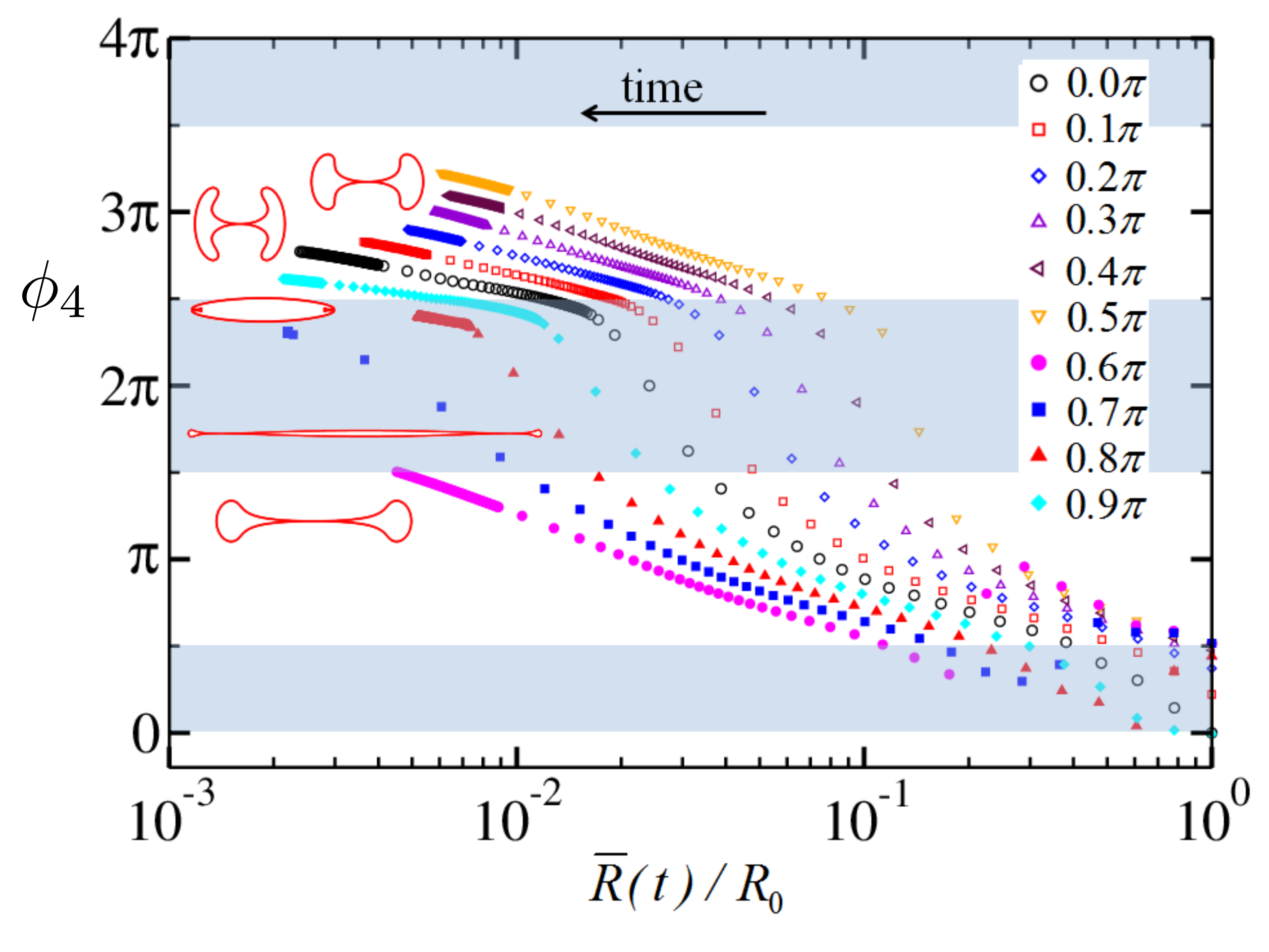}}
\caption{Evolution of the phase $\phi_4$ of mode $n=4$ over time for different initial values of $\Omega_2$ (see the legend). All the data sets (color online) display qualitatively the same behavior---the phase $\phi_4$ winds up more rapidly when modes $2$ and $4$ interfere destructively (shaded regions).  The phase speed slows when modes $2$ and $4$ interfere constructively. Cross-section shapes at the final moment for different initial states are displayed near the corresponding curves.}
\label{fig:interference}
\end{figure}

We plot the shape evolutions for different values of the initial phase $\Omega_2$ in figures \ref{fig:seq} and \ref{fig:slot}. Each column in figure \ref{fig:seq} corresponds to the shape evolution starting from a particular initial phase value.  Close-up's of shape evolutions that yield the more extreme values of $x_m/y_m$ are given in figure \ref{fig:slot}.  
Finally, following the suggestion from weakly nonlinear analysis that phase evolution controls shape evolution, we plot $\phi_4$ versus the rescaled average cross-section radius $\rbar/R_0$ in Fig~15 for several runs with different initial values of $\phi_2$.  We have not plotted $\phi_2 (\rbar)$ because the phase of the $n=2$ mode deviates only slightly from the linear stability prediction in all the runs and always has the same qualitative behavior as that shown in figure \ref{fig:mode2and4compare} (b). To make it easier to correlate the shape evolution data in figure \ref{fig:seq} with the phase evolution data in figure \ref{fig:interference}, we labeled a few curves with the final cross-section shapes and marked in shadow the interval where mode $2$ experiences destructive interference with mode $4$.  As with linear waves such as ripples over the surface of a pond, it is possible to read the complex pattern of shapes displayed in figure \ref{fig:seq} and \ref{fig:slot} in a simple way once the phase behaviors of the different  wave modes present (figure \ref{fig:interference})  are known.

Taken together, the shape evolution and the phase evolution clearly show that the nonlinear resonance between the initial perturbation and the higher harmonic modes it generates is responsible for the rich and sometimes counter-intuitive pinch-off dynamics exhibited by a distorted bubble.  Time-intervals during which modes $2$ and $4$ are in destructive interference are associated with a progression towards a cusp-like mode pinch-off, characterized by elongation of the cross-section shape and, more generally, the development of sharply curved regions along the interface.  In contrast, time-intervals during which modes $2$ and $4$ constructively interfere are correlated with evolution towards a coalescence-like mode of pinch-off, one where distant sections of the interface osculate. 

For small values of  $\Omega_2$, the $n=2$ and $n=4$ shape oscillations interfere constructively in the final stage of break-up (figure \ref{fig:interference}).  The two modes go through an interval of destructive interference initially, but this has a negligible effect on the shape evolution because the amplitude of the $n=4$ mode in the first stage is too small.  Thus the break-up is dominated by the effect of constructive interference.  The final aspect ratio is small and changes only slightly as $\Omega_2$ changes.  In this regime, the bubble pinch-off dynamics evolves into a coalescence mode break-up (see figure \ref{fig:seq}, first column). 

The second and third columns in figure \ref{fig:seq} display shape sequences produced by initial phase values slightly smaller than those of the peak in the aspect ratio $x_m/y_m$ curve in figure \ref{fig:aspectratioCompare}. Figure \ref{fig:slot}(a) gives a more detailed view of the final stage for the $\Omega_2 = 0.691 \pi$ run.  The $n=2$ mode interferes destructively with mode $4$ in this interval of $\Omega_2$ values, thereby causing the break-up to evolve into a double point contact. The simulation shows the same qualitative behavior, except that the bubble cross-section in the full simulation is far more elongated than the prediction of the weakly nonlinear analysis. This elongation is unique to $n=2$ perturbation. Initial states perturbed by higher $n$ modes develop sharp tips but retain an $O(1)$ aspect ratio \citep{lai12}.

The $4$th column in figure \ref{fig:seq} displays the shape evolution produced by starting the $n=2$ vibration with an initial phase that is slightly  higher than the value for the high peak in figure \ref{fig:aspectratioCompare}. In this interval of $\Omega_2$ values, the weakly nonlinear analysis predicts a near-miss (figure \ref{fig:wklynonlinear} (c)), with the final topology change taking place after the $n=4$ mode has completed a half cycle of shape oscillation.  This puts the $n=4$ oscillation again in constructive interference with the $n=2$ vibration.  The full simulation does something rather different.  It does not complete another half cycle of $n=4$ oscillation after the near-miss. Instead of broadening back into an oblong, slot-like shape after evolving towards a double contact,  two narrow, re-entrant fingers of water form and intrude inwards from the end cap regions (e.g., figure \ref{fig:slot} (b)).   The tips of these fingers evolve towards a cornered profile, with the curvatures at the two side-corners increasing extremely rapidly as a function of time. The shape appears to evolve towards a curvature singularity: small regions along the interface become highly curved and under-resolved, halting the simulation. An evolution that ended because of the formation of a high-curvature region is indicated by a solid black circle in figure \ref{fig:aspectratioCompare}. In practice, other physical effects presently neglected in our model, specifically the viscous drag and compressibility associated with the gas flow, can become important \citep{lai12}. An accurate account of the outcome produced by the cusp-like mode of pinch-off therefore requires us to track the surface evolution when a compressible, viscous gas flow in the interior is coupled with the incompressible, inviscid exterior liquid flow. While this challenging question lies outside the scope of the present study, our results here are sufficient to indicate that higher harmonics are generated with sufficiently large amplitude by nonlinear resonance to completely alter the final pinch-off dynamics from that predicted by weakly nonlinear analysis. 

A further increase of the value of the initial phase $\Omega_2$ creates the most counter-intuitive pattern of shape changes.  Here, the cross-section develops finger-like intrusions which start out highly curved (e.g., figure \ref{fig:seq}, second-to-last row in column (5) and 4th row in column (6)). Yet these sharp features do not continue to sharpen, even though the evolution becomes more nonlinear as break-up proceeds.  Instead the intrusions broaden, creating a smooth, $4$-fold contact (column (5) in figure \ref{fig:seq}). When the effect of the $n=2$ mode dominates that of the $n=4$ mode, the final pinch-off is a smooth single-point contact (column (6) in figure \ref{fig:seq}).  How this happens can be understood simply by looking at the phase evolution in figure \ref{fig:interference}.  The $4$-fold contact emerges from a shape evolution that has modes $2$ and $4$ in destructive interference throughout most of the final stage of pinch-off. However, in the very final moments, the $n=4$ shape oscillation completes a reversal.  This brings the two vibrations into constructive interference and a coalescence-like mode of pinch-off obtains.  This initial sharpening followed by subsequence broadening is essentially a counter-part of the double-contact formation sequence in column (3) of figure \ref{fig:seq}.  The double-contact sequence begins with $n=2$ and $4$ in constructive interference and ends with the two shape oscillations just at the verge of destructive interference. With the attendant effect of many higher harmonics being excited, this destructive interference drives the shape towards a cusp-like mode of pinch-off by causing the cross-section to elongate.  In  contrast, the onset of constructive interference causes the amplitudes of the higher harmonics to diminish, as a result shunting the evolution off from its original progression towards a cusp-like break-up and re-directing it towards a coalescence-like break-up. 

An additional increase in $\Omega_2$ restores the single-contact coalescence-mode of pinch-off (e.g., column (7) in figure \ref{fig:seq}), now between the left and right sides rather than the top and bottom sides of the interface. This shape sequence is simple and intuitive, because the two modes remain in constructive interference throughout so that they simply reinforce each other's effects. Because the break-up dynamics preserves the initial spatial symmetry, the shape evolution depicted in the last column for $\Omega_2 = 1.4 \pi$ in figure \ref{fig:seq} reproduces exactly the evolution in column (1) of figure \ref{fig:seq} for $\Omega_2 = 0.4 \pi$, except for the change in the spatial orientation by $90^\circ$. 


To recap, the full simulation results confirm the scenario from weakly nonlinear analysis; the dramatic and counter-intuitive changes in the cross-section profiles associated with underwater bubble break-up are inevitable consequences of phase dynamics and are largely insensitive to the details of shape evolution.  
The phase evolution of the fundamental mode, $\phi_2$, is basically insensitive to all other aspects of the pinch-off dynamics (figure \ref{fig:mode2and4compare} (b) and (d)).  It simply winds up linearly as a function of $\ln(R_0/\rbar)$.  The time-evolution of $\phi_4$ (the phase of the first higher harmonic generated) is only slightly more complex. It is insensitive to the higher modes and only sensitive to the orientation of the symmetry axis defined by the fundamental mode $n=2$, winding up more rapidly when the two modes are in destructive interference than when the two modes are in constructive interference (figure \ref{fig:mode2and4compare} (b) and (d), and figure \ref{fig:interference}).  
This simple behavior in the phase evolution is responsible for the rather counter-intuitive behaviors in the shape evolution. 
A switch from destructive interference between modes $2$ and $4$ to constructive interference re-directs the pinch-off dynamics from evolution towards cusp-like break-up to one towards coalescence-like break-up. Similarly, a switch from constructive to destructive interference diverts the shape evolution from progression towards a coalescence-like break-up to one towards a cusp-like break-up. Because the phases of the shape vibrations wind up in the same fashion regardless of how closely the shape has approached a singular shape or a topology change, the pinch-off dynamics can evolve almost up to the point of contact in a coalescence-mode break-up before apparently changing its mind entirely and evolving towards a cusp-like profile instead. This switch between constructive and destructive interference in the phase dynamics is responsible for the apparently counter-intuitive shape evolution displayed in the full simulation: even when the cross-section shape approaches closely to a singular shape or a topology change, the dynamics can halt itself and display a different behavior.

\subsection{Concavity oscillations in boundary integral simulation \& X-ray imaging}\label{subsec:nump2}
We next re-examine the high-speed X-ray imaging results in light of our simulations.  Previous simulation, as well as weakly nonlinear analysis and boundary integral simulations described here, all agree that a coalescence-mode pinch-off dynamics, one in which the bubble neck evolves into a thin air sheet (figure \ref{fig:xraycoalescence}), is a common outcome.  These correspond to initial $n=2$ mode perturbations that evolve into neck cross-sections that have high aspect ratios at the final moment.  We therefore focus on the peculiar concavity oscillations observed experimentally (figure \ref{fig:xraymemory}).  

In connecting the 2D boundary integral simulations with the experiments, we take advantage of the fact that the evolutions of the bubble cross-sections at different heights are decoupled at leading order.  We simply performed a series of two-dimensional cross-section shape oscillations, one appropriate for each vertical height, and interpolated these together to generate side views of the neck evolution.  This procedure is essentially the same as that employed by Schmidt et al.~\citep{schmidt09} to compare calculated and simulated results for cylindrically-symmetric evolutions.  There are two key assumptions. (i) The initial perturbations have the same relative amplitudes, $A_2/R_0$ and $A_4/R_0$, along the entire length of the bubble neck. (ii) Because cylindrically-symmetric pinch-off results in a vertical neck profile that asymptotes towards a hyperbola $R^2(z)=c^2z^2+R_{\rm min}^2$ (where $R_{\rm min}$ is the minimum neck radius), we fit a hyperbola to $R(z)$, the averaged neck radius as a function of vertical distance, and use that as the initial value for the average pinch-off dynamics.  

\begin{figure}
\centerline{\includegraphics[width=12cm]{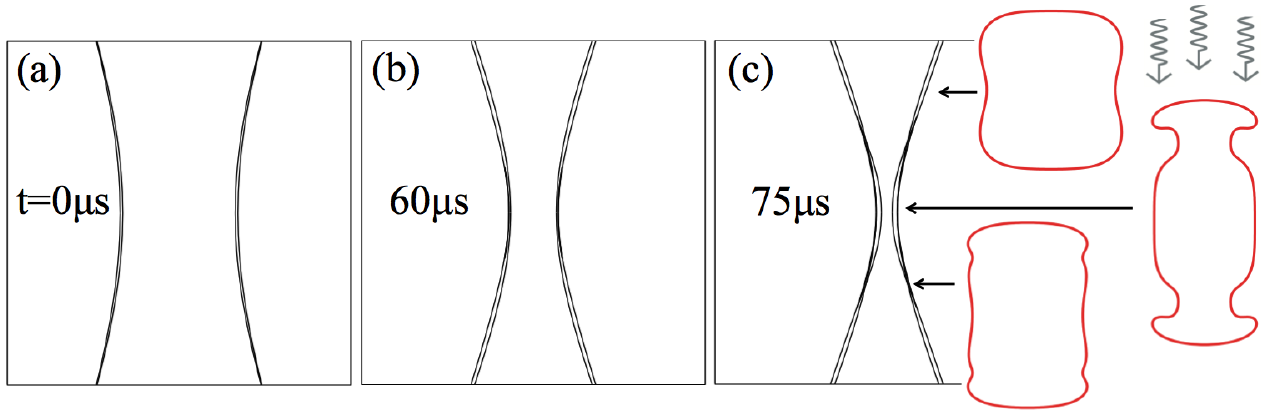}}
\caption{Simulated underwater bubble pinch-off dynamics reproduces the peculiar concavity oscillation observed in high-speed X-ray imaging (figure \ref{fig:xraymemory}). The initial perturbation is a mixture of $n=2$ and $n=4$ shape oscillations.  (a) Initially, the bubble neck shape shows little evidence of azimuthal perturbations. (b) About $60$ $\mu$s afterwards, the effect of the $n=2$ mode becomes evident.  The cross-section has two regions whose directions parallel the X-ray beam, thus giving rise to $2$ fringes in the profile view. (c) About $75$ $\mu$s later, the break-up dynamics is dominated by the perturbation modes.  These have evolved to create highly corrugated cross-section shapes which vary significantly along the vertical direction. }
\label{fig:sim_neck}
\end{figure}

Simulations starting with a variety of initial neck shapes, all perturbed by a single $n=2$ mode perturbation, fail to reproduce the concavity oscillation observed in the X-ray experiment.  In contrast, a more complex initial perturbation, one comprised of both $n=2$ and $n=4$ modes, easily gives rise to the concavity oscillation observed.  This choice of initial modes is consistent with the experimental geometry, where an air bubble is released into a square cross-sectioned container from a slot-shaped nozzle. Figure \ref{fig:sim_neck} depicts one representative outcome, obtained using the initial state ($A_2/R_0=0.1$, $\Omega_2=0.4\pi$) and ($A_4/R_0=0.05$, $\Omega_4=0.7\pi$). As observed in the experiment (figure \ref{fig:xraymemory}), vertical fringes develop as the neck evolves. These  correspond to regions where the interfaces are parallel to the X-ray beam direction.  For the final moment, we also plot the simulated cross-sections at different heights.  The signatures of both $n=2$ and $n=4$ oscillation are evident.  This is not a unique assignment of the amplitudes and phase - other values of $A_2$, $\Omega_2$, $A_4$ and $\Omega_4$ generate comparably good agreement with the experimental images. 

While the presence of both modes in this case prevent a definitive assignment of the observed concavity oscillation with a unique initial state, our survey of initial states needed to reproduce such observations makes a strong case that these oscillations require the presence of multiple perturbation modes in the initial neck shape.  These oscillations therefore provide a visible and dramatic evidence of nonlinear resonance in underwater pinch-off dynamics. 

\section{Discussion}\label{sec:discussion}
Several questions are naturally raised by our results. The relation between the severely elongated cross-section shapes predicted by our simulation and the experiment is unclear.  While the final break-up dynamics in the experiment does appear to be more sensitive to the initial condition in some ranges than others, we have found it difficult to reproduce this sensitivity reliably.  Viscous effects in the gas can transform a coalescence-mode break-up into a  double-point contact, making it difficult to correlate solutions of our idealized model with the experiment.  This is complicated further by the fact that the natural parameter, the phase of the vibration imposed at the initial moment, is difficult to control in the experiment.  Cavity collapse experiments pull a distorted solid disk downwards into a body of water to create an initial shape with prescribed corrugations \citep{enriquez12}.  In this case, the downward flow creates a level of background disturbances that make the initial phase value difficult to reproduce from run to run.  The bubble release experiments conducted here are able to reproduce initial phase values reliably but we have found it difficult to vary the value of the initial phase freely.  

Another set of questions relates to the result that the curvature can become very large in shape evolutions towards cusp-like modes of pinch-off.  While surface tension effects are unimportant for coalescence-mode break-up with $O(1)$ aspect ratio, they may become significant for the sharply curved, severely elongated cross-section shapes displayed in figure \ref{fig:seq}.  It is also not clear whether the tip curvature attains a maximum value or diverges, even in the idealized situation analyzed here where surface tension is absent and there is no dissipation.  This question is being addressed in a separate study \citep{lai12}.   

A third set of questions consider how our results, which focus on the simple case of perturbation by a single Fourier mode, change in more general situations.  As already discussed in the paper, the form of the governing equations (equations (\ref{Reqn}) and (\ref{Veqn})) dictates that initial conditions with a single Fourier mode can only generate higher harmonics. For example, a mode $n=20$ perturbation generates $n=40$... but not lower modes such as mode $n=10$. We therefore expect that an initial state perturbed by a single, different Fourier mode will follow the same scenario as that for $n=2$. Within our model, we looked at several examples when the initial state is perturbed by a single $n=3$ mode and similar evolutions as in figure \ref{fig:interference} are observed for mode $n=6$ (the lowest higher harmonic generated by mode $n=3$), i.e. the phase of mode $n=6$ winds up more rapidly when it is in destructive interference with $n=3$ than that when it is in constructive interference with $n=3$, and the final outcomes with smooth contact coincide with the constructive interference between modes $n=3$ and $6$ while the cusp-like break-ups coincide with the destructive interference. Although we expect a similar behavior for mode $2n$ when initial state is perturbed by a single mode $n>2$, a systematic numerical check requires a different simulation scheme with better resolution and the ability to track Fourier modes with high mode numbers. In practice, shapes with larger $n$ modes will be modified more strongly by viscous and surface tension effects during the initial stage.  We expect perturbations by Fourier modes larger than $20$ to be largely irrelevant for underwater bubble break-up, since surface tension effects will very quickly smooth out the shape.  They can be relevant in the collapse of large air cavities, as demonstrated by Enriquez \emph{et al.}'s experiment \citep{enriquez12}.  It is less clear what obtains if two incommensurate Fourier modes are imposed at the initial moment since interaction between the two modes should then generate all possible higher harmonic modes, instead of a subset which has the same spatial symmetry as the original single-mode Fourier perturbation.  

Along a different front, there is clear experimental evidence that an $n=1$ perturbation, i.e. a tilt of the neck centerline, has a strong effect on the break-up and subsequent jetting (figure \ref{fig:xrayjet}).  Since a tilting of the neck centerline and the jetting inevitably create a weak vertical flow that couples the evolutions of the cross-sections at different heights, modeling these effects quantitatively requires that we modify the strictly two-dimensional model used here to allow momentum transfer at different vertical heights.  Prior studies have examined only cylindrically-symmetric dynamics with a straight centerline and found it stable against the perturbative effects of vertical flow \citep{eggers07,gekle09b,herbst11}.  It will be interesting to extend these results to account for the tilt-induced dynamics observed here. 

We also know little about how the scenario uncovered in this study changes when the  initial amplitude of the perturbation is altered.  Typically, increasing the amplitude makes the nonlinearities stronger. However, in our situation, a larger perturbation also causes the bubble to pinch off earlier, thus taking away time for the nonlinear resonance to create and amplify higher harmonic modes.  It is not obvious whether the increased strength of the interaction associated with a larger initial amplitude can compensate for the shorter amount of time available for break-up.  Results from this study indicate that the behavior will vary, depending on the relative phase difference between the higher harmonics and the initial perturbation.  

\section{Conclusion}\label{sec:conclusion}
In conclusion, we have presented results from high-speed X-ray imaging, weakly nonlinear analysis and boundary integral simulations, showing that the final break-up dynamics exhibits two qualitatively different types of behaviors: a coalescence-like mode of break-up, and a cusp-like mode of break-up that involves the formation of sharp tips. 
Weakly nonlinear analysis accounting for only the initial perturbation and the first higher harmonic generated is able to predict that the break-up dynamics changes abruptly when the two modes interfere destructively, rather than constructively, as break-up approaches.  However, it fails to describe the resultant form of  dynamics because the destructive interference triggers the amplification of many higher harmonic modes. 

More broadly speaking, our results show not only that a memory of the initial perturbation controls the final topology  change in underwater bubble break-up but also that nonlinear resonance among the different shape oscillations causes the consequences of this memory to be highly variable. The basic picture that emerges is that coalescence-mode break-up, which is associated with the constructive interference between the initial perturbation mode and the first higher harmonic generated and hence is a simple extrapolation of the memory-preserving linear stability behavior, retains a relatively faithful recollection of the initial condition.  In contrast, the host of higher harmonics created and amplified at the onset of the destructive interference between the initial mode and the first higher harmonic generated appears to create a kind of false memory. When these two modes are in destructive interference, various complex break-ups are observed. These complex break-ups include the oblong final cross-section shape that obtains experimentally by releasing a bubble from a slot at high speed instead of a quasi-static protocol. They also include a cusp-like mode of break-up that is characterized by local regions whose radii of curvature are much smaller than the average cross-section radius. The switch between constructive and destructive interference in the phase dynamics dictates the shape evolution: a cross-section shape evolves towards one type of break-up can halt itself and starts evolving towards a different type of break-up.

\vspace*{0.2in}
\noindent{\bf Acknowledgements}

We thank Konstantin Turitsyn for first noticing the possibility of a new cusp-like break-up dynamics. Thomas Caswell, Michelle Driscoll, Joseph Paulsen and Ariana Strandburg-Peshkin assisted in carrying out the experiments.  Eric Brown critiqued a preliminary version of the manuscript.  We also acknowledge helpful conversations with Samuel Oberdick and  Elizabeth Hicks. Use of the Advanced Photon Source was supported by the U.S. Department of Energy, Office of Science, Office of Basic Energy Sciences, under Contract No. DE-AC02-06CH11357.  This research was supported by NSF-CBET 0730629 (W. W. Z.) and MRSEC DMR-0820054. 

\appendix
\section{Reformulation in terms of canonical ${\cal R}$, ${\cal V}$ variables}
By maintaining the analyticity of the complex velocity potential $\Psi(w,t)$ in the exterior of the unit circle, its real part, the velocity potential $\Phi(w, t)$, always satisfies Laplace's equation. The solution of $\Phi(w,t)$ then allows us to solve for the time evolutions of the surface and the velocity potential on the surface using boundary conditions (\ref{eqn:kinematicbc}) and (\ref{eqn:dynamicbc}), which is identical to solving for the time evolution of the mapping ${\cal R}(w, t)$ and ${\cal V}(w, t)$. The analyticity of $z(w,t)$ and $\Psi(w,t)$ requires us to analytically continue equations (\ref{eqn:kinematicbc}) and (\ref{eqn:dynamicbc}) before rewriting them in terms of ${\cal R}$ and ${\cal V}$ variables. For equation (\ref{eqn:kinematicbc}), this is achieved by adding a tangential velocity on the surface along with the normal component of the velocity to ensure the analyticity of the partial time derivative of the mapping $z(w,t)$ \citep{shraiman84} even though it would not change the surface shape. For equation (\ref{eqn:dynamicbc}), the analyticity of $\partial_t\Psi(w,t)$ is ensured by adding an appropriate imaginary part (corresponding to the partial time derivative of the stream function $\psi=Im(\Psi)$) to both sides of the equation, and $P(t)$ can be absorbed in the definition of $\Psi$. Therefore, from Shraiman and Bensimon \citep{shraiman84} equation (\ref{eqn:kinematicbc}) is rewritten as 
\begin{equation}\label{zteqn}
\partial_tz(w,t)=w\partial_wz\left[\frac{Re(w\partial_w\Psi)}{|w\partial_wz|^2}+iC\right]
\end{equation}
Equation (\ref{eqn:dynamicbc}) can be rewritten as
\begin{equation}\label{Psiteqn}
(\partial_t\Psi(z(w,t),t))_z=-\frac{1}{2}|\partial_z\Psi|^2+iD
\end{equation}
where $(\partial_t\Psi(z(w,t),t))_z$ means the partial derivative with respect to $t$ with $z$ fixed. Both $C$ and $D$ are real-valued functions of $w$. They are added as described above to make the right-hand sides of equations (\ref{zteqn}) and (\ref{Psiteqn}) analytic and their function forms are determined by the Poisson integral formula. In our case this formula states that if we know that the real part of a function analytic in the exterior of the unit circle $w=e^{i\alpha}$ (on the $w$-plane) can be written on the unit circle $w=e^{i\alpha}$ as
\begin{equation}\label{fexpansion}
f(\alpha)=a_0+\sum_{n=1}^{\infty}(a_ne^{in\alpha}+a_n^*e^{-in\alpha})
\end{equation}
then this analytical function takes the following form
\begin{equation}\label{Aexpansion}
{\cal A}\{f\}(w)=a_0 + 2\sum_{n=1}^{\infty}\frac{a_n^*}{w^{n}}
\end{equation}
where $\cal A$ is an integral operator applied to a real-valued function $f(w)$ defined on the unit circle $w = e^{i\alpha}$ on the $w$-plane, resulting in an analytical function ${\cal A}\{f\}(w)$ defined in the exterior of the unit circle such that $Re({\cal A}\{f\}(e^{i\alpha}))=f(e^{i\alpha})$. In an integral form, we have
\begin{equation}\label{Aintegral}
{\cal A}\{f\}(w)=\frac{1}{2\pi i}\oint_{|w'|=1}\frac{dw'}{w'}\frac{w+w'}{w-w'}f(w')
\end{equation}

\section{Numerical implementation of the boundary integral method}
Our numerical implementation of the boundary integral equation that governs the bubble evolution follows closely the method described by Pozrikidis \citep{pozrikidis}, with minor variations to accommodate the needs of our specific problem.  For the sake of completeness, we include a brief description of the implementation below (please see \citep{pozrikidis} for more details). We discretize the interface by $N+1$ nodal points $\bf{x_i}$ $(i = 1,...,N+1)$. The index $i$ increases counter-clockwise along the interface. $\bf{x_1}$ and $\bf{x_{N+1}}$ coincide with each other for a closed interface. $N$ boundary elements $e_k$ $(k=1, ... ,N)$ are constructed by connecting adjacent nodal points $\bf{x_k}$ and $\bf{x_{k+1}}$ using straight line segments and the position $\bf{x}$ along the interface is interpolated using piecewise linear interpolation between adjacent nodes. The potential $\Phi^e_{k}$ and normal velocity $u^e_{\perp,k}$ on each element $e_k$ are assumed to be constant. Instead of solving for the normal velocity at each nodal point directly, equation (\ref{integraleqn}) is used to solve for the normal velocity at the middle point $\bf{x^e_k}$ of each element $e_k$ (equivalently the normal velocity on that element), which helps suppress numerical instabilities. With all the asuumptions, equation (\ref{integraleqn}) now takes a discretized form (after re-arranging the terms)
\begin{equation}\label{discretizedeqn}
\sum_{k}\Big[\int_{{\bf x}\in e_k}G({\bf x^e_j}, {\bf x})ds({\bf x})\Big]u^e_{\perp,k}= \sum_{k}\Big[\int_{{\bf x}\in e_k}({\bf n}\cdot{\bf \nabla})G({\bf x^e_j}, {\bf x})ds({\bf x})\Big]\Phi({\bf x^e_k})-\frac{1}{2}\Phi({\bf x^e_j})
\end{equation}
where each integral in the summation is over one line element. Equation (\ref{discretizedeqn}) can be written in a more compact form with matrices 
\begin{equation}\label{matriceseqn}
\sum_k(SL)_{jk}u^e_{\perp,k}=\sum_k(DL)_{jk}\Phi^e_k - \frac{1}{2}\Phi^e_j
\end{equation}
where $N$-by-$N$ matrices $SL$ and $DL$ only depend on the interface shape. As in equation (\ref{greensfunction}), the Green's function $G({\bf x_0}, {\bf x})$ has a logarithmic divergence as $\bf{x}$ approaches $\bf{x_0}$. As a result, diagonal elements $(SL)_{kk}$ (or $(DL)_{kk}$) in $SL$ (or $DL$) are singular integrals since the point ${\bf x^e_k}$ where the potential and normal velocity are evaluated falls on the element $e_k$ where the line integral is performed. However, for these line elements, those integrals can be calculated analytically, which gives $(SL)_{kk}=-L^e_k(\ln(\frac{1}{2}L^e_k)-1.0)/(2\pi)$ and $(DL)_{kk}=0$ where $L^e_k = |\bf{x_{k+1}}-\bf{x_k}|$ is the length of the $k$th element. The non-singular integrals in the off-diagonal elements in $SL$ and $DL$ are computed using 20-point Gaussian quadrature. 

To start our simulation, the initial position $\bf{x_i}$ and potential $\Phi_i$ at each nodal point $\bf{x_i}$ are generated from prescribed specific function forms of ${\cal R}(w)$ and ${\cal V}(w)$. At each time step $t$, starting from the values of $\bf{x_i}$ and $\Phi_i$ at each nodal point, the potential $\Phi^e_k$ on each element $e_k$ is obtaned by taking the average of the potentials at the end points $\bf{x_k}$ and $\bf{x_{k+1}}$, i.e. $\Phi^e_k=\frac{1}{2}(\Phi_k+\Phi_{k+1})$. Then the normal velocity $u^e_{\perp,k}$ at the middle point $\bf{x^e_k}$ of $e_k$ is solved by solving equation (\ref{matriceseqn}) using the Gaussian elimination as $u^e_{\perp}= SL^{-1}(DL-\frac{1}{2}I)\Phi^e$, where $u^e_{\perp}=(u^e_{\perp,1},\ldots,u^e_{\perp,N})^T$ and $\Phi^e=(\Phi^e_1,\ldots,\Phi^e_N)^T$ are two column vectors and $I$ is the $N$-by-$N$ identity matrix. The normal velocity $u_{\perp,i}$ at each nodal point is obtained by taking the average of the normal velocities on its adjacent elements $u^e_{\perp,i}$ and $u^e_{\perp,i-1}$ (for a closed interface, when $i=1$, the subscript $i-1$ is understood as the value $N$), weighted by the length of the elements $L^e_i$ and $L^e_{i-1}$ as $u_{\perp,i}=(u^e_{\perp,i}L^e_{i-1}+u^e_{\perp,i-1}L^e_i)/(L^e_i+L^e_{i-1})$. For re-meshing purpose, the tangential component of the velocity $u_{\parallel,k}$ at each nodal point is also calculated by differentiating the velocity potential $\Phi$ along the interface numerically, i.e., $u_{\parallel,k}=\left(\frac{\Phi_{k+1}-\Phi_k}{L^e_{k}}L^e_{k-1}+\frac{\Phi_k-\Phi_{k-1}}{L^e_{k-1}}L^e_k\right)/\left(L^e_{k}+L^e_{k-1}\right)$. After that, we update the $N+1$ nodal points $\bf{x_i}$ and the potential $\Phi_i$ to time $t+\Delta t$ using the total velocity ${\bf u}=u_\perp{\bf n}+u_\parallel{\bf t}$ according to equations (\ref{eqn:kinematicbc}) and (\ref{eqn:dynamicbc}), where $\bf{n}$ and $\bf{t}$ are unit normal and tangent vectors on the surface respectively. The value of $P(t)$ in equation (\ref{eqn:dynamicbc}) will shift the potential at time $t+\Delta t$ on the surface by a constant which affects the normal velocity according to equation (\ref{integraleqn}), and the value of $P(t)$ is determined to ensure a prescribed areal flux, which is constant in our simulation. The new time step will repeat the above steps starting from the updated $\bf{x_i}$ and $\Phi_i$ at each nodal point. 

\bibliographystyle{jfm}
\bibliography{phaseSweep}
\end{document}